%% file: neurips_2026.tex
\pdfoutput=1
\documentclass{article}

\PassOptionsToPackage{numbers,compress}{natbib}

\usepackage[main, final]{neurips_2026}

\usepackage[utf8]{inputenc} 
\usepackage[T1]{fontenc}    
\usepackage{hyperref}       
\usepackage{url}            
\usepackage{booktabs}       
\usepackage{amsfonts}       
\usepackage{amsmath}        
\usepackage{nicefrac}       
\usepackage{microtype}      
\usepackage{xcolor}         
\usepackage{graphicx}       
\usepackage{multirow}       

\newcommand{\injname}[0]{MCP-Injector}

\title{Tool-Calling LLM Agents Under Realistic MCP Faults: Scale Does Not Purchase Fault Tolerance}

\author{%
  Antonio Ken Iannillo\thanks{Corresponding author.} \\
  SnT, University of Luxembourg \\
  \texttt{antonioken.iannillo@uni.lu} \\
  \And
  Joshua Segun Owotogbe \\
  Tilburg University \\
  \And
  Roberto Natella \\
  Gran Sasso Science Institute \\
  \AND
  Francesco Avallone \\
  University of Naples Federico II \\
  \And
  Kristina Kudryavtseva \\
  SnT, University of Luxembourg \\
  \And
  Indika Kumara \\
  Tilburg University \\
}

\begin{document}

\maketitle

\begin{abstract}
  LLM agents increasingly rely on external tools exposed via the Model Context Protocol (MCP), yet current benchmarks assume tools respond correctly and on time, leaving agent resilience to realistic tool failures untested.
  We introduce \injname{}, a drop-in, protocol-aware fault-injection middleware that interposes on the MCP tool-call path with controlled persistence (permanent, transient, intermittent) and seedable schedules.
  Its fault model is calibrated to the observed incident distribution (failure mode $\times$ persistence) of the public MCP ecosystem, mined from 57,240 issue and pull-request threads across 1,774 open-source MCP repositories.
  In a 1,350-cell experiment crossing two backbone sizes (20B, 120B), three agent stacks (custom multi-round plan-and-execute, OpenAI Agents SDK, LangChain), and three fault injection conditions, we find that faults drawn from this permanent-heavy stress profile cause severe quality degradation (task-completion score falls by 60\%, from 8.08 to 3.23 out of 10), and that scale does not purchase fault tolerance for the two gpt-oss backbones we test: the 120B retains no larger share of its fault-free score under faults (resilience-ratio model, $p = 0.96$), and its completion rate under faults is, if anything, lower (91.1\% vs.\ 95.6\%; difference 4.4 points, 95\% CI $[-0.2, 9.3]$).
  Agent stacks differ in how well they absorb faults, which suggests that framework-level error handling matters more than model scale.
  These results demonstrate that capability benchmarks alone can mask operational brittleness, motivating protocol-aware fault injection as a complementary evaluation dimension for tool-calling agents.
\end{abstract}

\section{Introduction}
\label{sec:introduction}

LLM agents increasingly act through external tools \cite{gim2025pie, luo2023self} exposed via the Model Context Protocol (MCP), a vendor-neutral, structured client-server interface for tool invocation \cite{11105280,11205601,11213046}. This makes agent quality a \emph{system} question: end-to-end behavior depends on the MCP channel, server implementations, schemas, and the operational health of the tool ecosystem. In production, MCP-facing agents inherit failure modes familiar from distributed services \cite{1335465, 10206870}, yet current benchmarks evaluate agents assuming tools respond correctly and on time \cite{yao2024tau,wang2024gta,wang_mcp-bench_2025,mialon_gaia_2023}. Without protocol-aware fault injection, the community lacks a controlled method to answer: \emph{Which agent designs degrade gracefully? Does scaling the backbone reduce or increase operational brittleness?}

Because MCP standardizes \emph{structured} tool calls, failures can be semantic: a well-formed but wrong value, a schema violation, a silent hang, or a retry cascade. These behaviors align with classic content and timing failure classes \cite{1335465} but must be expressed at the MCP interface to exercise agent logic that interprets outputs, retries, and composes answers. Infrastructure-level chaos engineering \cite{7436642} operates below this semantic layer, and benchmarks do not control fault persistence or provide calibrated profiles, leaving a gap for \emph{benchmark-compatible, MCP-semantic} fault injection. We introduce \injname{}, a drop-in middleware that interposes on the MCP client-server path and injects faults with controlled persistence (permanent, transient, intermittent), with a fault model calibrated to the observed incident distribution of the public MCP ecosystem (57,240 issue and pull-request threads across 1,774 open-source MCP repositories) and packaged as a seedable injection schedule applicable to any MCP-based agent without modifying the agent or the tools.

\textbf{Contributions.} (1)~An MCP fault model calibrated to the observed incident distribution (mode $\times$ persistence) of the public MCP ecosystem, mined from 57,240 threads across 1,774 open-source repositories, so that the mix of injected faults follows reported incidents rather than hand-crafted scenarios. (2)~A 1,350-cell full-factorial study (three stacks, two backbone scales, three fault conditions, 25 tasks, three seeds) showing that scale does not purchase fault tolerance for the two gpt-oss backbones tested (the larger backbone retains no larger share of its fault-free score under calibrated faults and completes descriptively fewer tasks) and suggestive evidence that framework-level error handling matters more than model scale. (3)~\injname{} itself, a reproducible, stack-agnostic, drop-in middleware released with full artifacts.

\section{Background and Definitions}
\label{sec:background}

Tool-using LLM agents are implemented as an \emph{MCP client} embedded in an agent runtime, connected to one or more \emph{MCP servers} that expose executable \emph{tools}. We treat the \textbf{tool-call exchange} (the structured invocation request and its result or error) as the key service interface and call it the \textbf{MCP tool-call boundary}. Our experiments and injector operate strictly at this boundary; other MCP capabilities (prompts, resources, streaming) are out of scope, since tool calls are the dominant actuator by which agents affect the external world. We adopt a \textbf{non-adversarial threat model}: injected faults represent commonplace production problems (misconfiguration, transient unavailability, schema drift, malformed outputs), not an intelligent attacker; security attacks, model-internal faults, and unrelated application-layer issues are excluded.

\paragraph{Dependability terminology.}
We use standard dependability terminology \cite{1335465} instantiated at the MCP tool-call boundary. A \textbf{service failure} is a deviation of observed tool-call behavior from what the client expects (e.g., an error instead of a result, excessive latency, a schema-violating result); a \textbf{fault} is its underlying cause (a buggy implementation, misconfiguration, or dependency outage). Because systems compose, a failure of the MCP server is an \emph{external fault} from the client's perspective. Our injector injects such external faults at the \emph{interface} between the agent and the MCP server, rather than within server internals, for more controlled and efficient experiments as typical of fault injection studies \cite{natella2016assessing}.

\paragraph{Failure taxonomy and persistence.}
At the MCP boundary, interface failures divide into two classes \cite{1335465}: \textbf{timing failures} (results not delivered within the expected time budget, excessively delayed, or never delivered) and \textbf{content failures} (results that arrive but are wrong or unusable: incorrect values, malformed payloads, schema/type violations). Beyond \emph{what} fails, dependability evaluation must model \emph{how} failures evolve. We treat \textbf{persistence} as the central temporal dimension: \textbf{permanent} failures recur on every invocation until external corrective action; \textbf{transient} failures are short-lived and resolve without intervention; \textbf{intermittent} failures alternate between faulty and correct behavior in temporally correlated bursts. Our injector operationalizes these persistence modes as injection schedules over the tool-call boundary.

\section{\injname}
\label{sec:design}

\injname{}\footnote{Artifact (code, data, analysis pipeline): \url{https://gitlab.com/uniluxembourg/snt/sedan/papers/mcp-injector}, archived at \url{https://doi.org/10.5281/zenodo.23573765}.} is a protocol-aware fault-injection middleware for tool-calling LLM agents that communicate over MCP. It is \emph{drop-in} (no modifications to the agent runtime, MCP client, or servers), \emph{reproducible} (seeded injection schedules), \emph{low-interference} (minimal perturbation when faults are disabled), and \emph{stack-agnostic}.

\subsection{Architecture}
\label{sec:design-architecture}

\begin{figure}[t]
    \centering
    \includegraphics[width=\linewidth]{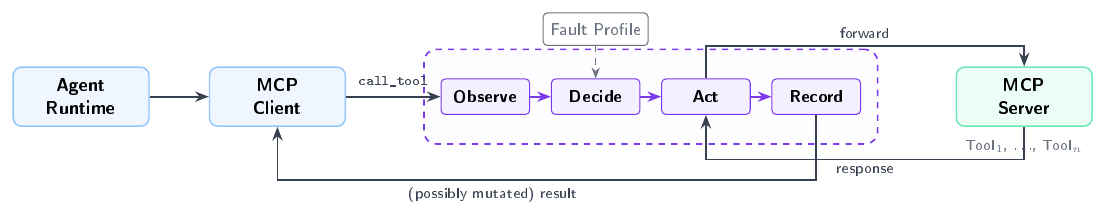}
    \caption{\injname{} architecture. The middleware transparently interposes on the MCP tool-call path. Each invocation passes through four stages: \textbf{Observe} (read tool name, arguments, and context), \textbf{Decide} (select an operator based on the fault profile, persistence state, and seeded randomness), \textbf{Act} (forward the call to the real server and optionally mutate the response), and \textbf{Record} (emit a structured log entry for post-hoc analysis).}
    \label{fig:architecture}
\end{figure}

\injname{} is implemented as proxy-style middleware on the tool-invocation path (Figure~\ref{fig:architecture}). It intercepts the \texttt{call\_tool} hook exposed by MCP client libraries, acting as an MCP server toward the upstream agent and as an MCP client toward the downstream real server(s), and can be toggled on or off via a single configuration flag. Request-side interception handles timing faults (delays, silent drops) and explicit errors that can be synthesized without contacting the server; response-side interception handles content faults by mutating the real server's output before it reaches the agent.

\subsection{Fault API}
\label{sec:design-fault-api}

\injname{} exposes faults through a configuration-driven API that separates three orthogonal concerns: \emph{what} failure to emulate (operator), \emph{how long} it persists (schedule), and \emph{where} it applies (scope).

\paragraph{Operators.}
Table~\ref{tab:operators} lists the six fault operators plus one passthrough, each corresponding to an empirically observed MCP failure mode (Section~\ref{sec:methodology}). Operators either short-circuit the request (explicit error, silent drop), delay forwarding, or mutate the response content.

\begin{table}[t]
\centering
\caption{Fault operators implemented by \injname{}. Content operators act on the server's response; timing operators act before or instead of forwarding.}
\label{tab:operators}
\small
\begin{tabular}{@{}llll@{}}
\toprule
\textbf{ID} & \textbf{Operator} & \textbf{Class} & \textbf{Effect} \\
\midrule
OP0 & Null injector    & n/a      & Passthrough (no mutation) \\
OP1 & Minor mistake    & Content  & Random character replacement \\
OP2 & Incorrect output & Content  & Word-level substitution \\
OP3 & Invalid output   & Content  & Type coercion (e.g., str$\to$int) \\
OP4 & Explicit error   & Content   & Returns MCP error message \\
OP5 & Delayed response & Timing   & Artificial sleep before return \\
OP6 & Silent drop      & Timing   & Indefinite hang (no response) \\
\bottomrule
\end{tabular}
\end{table}

\paragraph{Persistence schedules.}
\injname{} realizes the three persistence regimes of Section~\ref{sec:background} as seeded schedules: \textbf{Permanent} fires on every subsequent matching call once activated, \textbf{Transient} fires on a single call with a configured activation probability and then resets, and \textbf{Intermittent} holds the drawn entry for a random number of calls (released with a per-call deactivation probability) and fires on each held call with an activation probability.

A fault profile may combine multiple (operator, persistence) entries with configured selection weights, which \injname{} samples to produce realistic failure mixtures under seeded fault schedules: injection decisions are drawn from a seeded pseudo-random stream, so they are deterministic given the seed and the agent's tool-call history, and every fired fault is logged. End-to-end runs are not bit-reproducible, because local serving and the judge are stochastic (Appendix~\ref{app:profile-derivation}).

\section{Fault Profile Methodology}
\label{sec:methodology}

\begin{figure}[t]
    \centering
    \includegraphics[width=\linewidth]{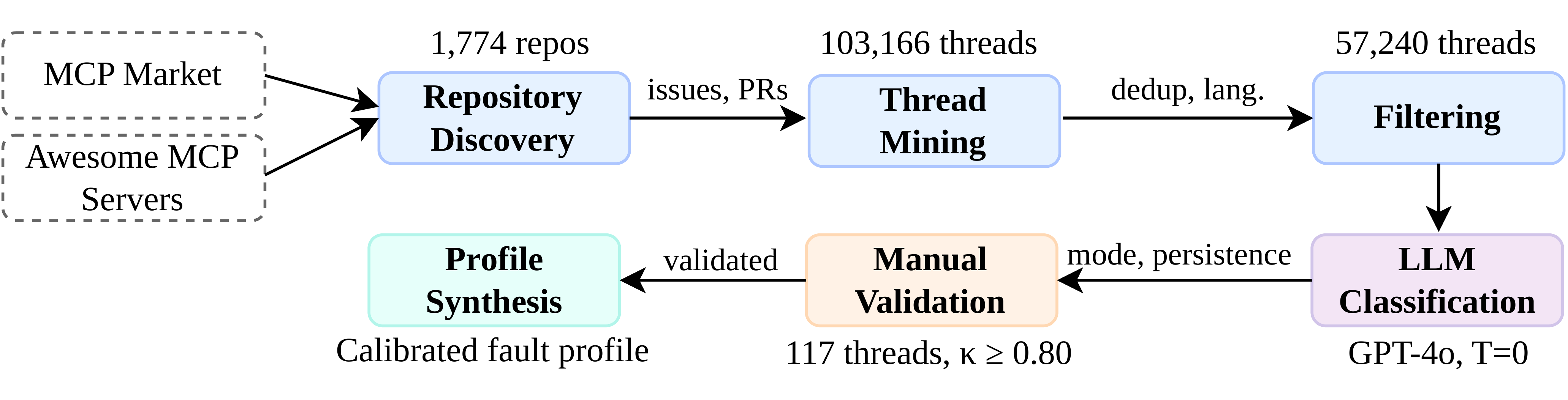}
    \caption{Fault-profile calibration pipeline: from repository discovery to a seedable injection profile. Numbers indicate corpus size after each stage.}
    \label{fig:mining_pipeline}
\end{figure}

We derive a fault profile, calibrated to the observed incident distribution of the public MCP ecosystem, that specifies \emph{what} MCP-visible failure manifestations to inject and \emph{how} they evolve over time.
Rather than relying on ad hoc assumptions, we mine public MCP tool-server repositories and classify observed incidents using standard dependability terminology~\cite{1335465}, specialized to the MCP tool-call boundary (Section~\ref{sec:background}).
Figure~\ref{fig:mining_pipeline} summarizes the six-stage pipeline.

\paragraph{Fault taxonomy.}
Within the content and timing classes of Section~\ref{sec:background}, we distinguish six mutually exclusive \textbf{failure modes}. Content failures: \textit{minor mistake} (small deviation, structurally valid), \textit{incorrect output} (semantically wrong, structurally valid), \textit{invalid output} (schema/type violation), and \textit{explicit error} (structured error in place of a result). Timing failures: \textit{delayed response} and \textit{silent drop}. Together with the three persistence regimes of Section~\ref{sec:background}, these modes define the (mode, persistence) target space for calibration.

\paragraph{Repository discovery and thread mining.}
We assembled \textbf{1,774} active public MCP tool-server repositories (collected September 2025) by combining two public registries, MCP Market\footnote{\url{https://mcpmarket.com/server}} and Awesome MCP Servers\footnote{\url{https://github.com/punkpeye/awesome-mcp-servers}}, deduplicating, and removing inaccessible ones. For each repository we mined up to 200 recent issues and pull requests via the GitHub REST API (103k issues; 64k PRs; 158k comments; 258k commits; 71k reviews; 65k review comments) and constructed per-thread narratives by concatenating all artifacts in chronological order, truncating threads above 50k characters to prioritize early failure descriptions. This stage yielded \textbf{103,166} consolidated threads.

\paragraph{Filtering.}
A keyword-based heuristic partitioned threads into failures, non-failures, and uncertain cases following prior issue-mining work~\cite{seo2014programmers}. After removing 1,374 cross-link duplicates, excluding build-related threads (3,013) and non-English threads (2,668) across automated and manual passes, the analysis set contained \textbf{57,240} English-language classified threads.

\paragraph{LLM-based classification and manual validation.}
We used gpt-4o-2024-08-06 (temperature 0, fixed JSON schema, $\sim$\$800 in API cost) to classify each thread into (failure mode, persistence) pairs and extract mode-specific parameters (error subtypes, data-type patterns, delay magnitudes). To assess reliability, two senior researchers independently annotated a stratified sample of \textbf{117} threads (five per non-empty stratum) with a third adjudicating disagreements; agreement was high (Cohen's $\kappa = 0.93$ for failure mode, $0.80$ for persistence~\cite{landis1977measurement}).

\paragraph{Profile synthesis.}
Aggregating classifier outputs yields the empirical distribution in Table~\ref{tab:joint_mode_persistence}. Permanent failures predominate, with \textsc{explicit\_error} the largest share (10,393 of 14,829 classified failures); other content failures (\textsc{invalid\_output}, \textsc{incorrect\_output}, \textsc{minor\_mistake}) are less frequent but operationally significant due to silent propagation risk; timing failures (\textsc{delayed\_response}, \textsc{silent\_drop}) span permanent and intermittent persistence. We package these distributions as a versioned, seedable fault profile consumed by \injname{} (Section~\ref{sec:design}).

\paragraph{What is calibrated.}
The profile matches the relative frequency of (mode, persistence) cells among mined failure reports, not operational failure rates. Data-derived: the 18 selection weights ($\mathrm{count}/N$, $N = 13{,}824$) and the six INTERMITTENT activation probabilities (Equation~\eqref{eq:activation_p_derivation}). Hand-set: $d = 0.3$, the 30\,s delay, the OP1/OP2 perturbation intensities, and TRANSIENT activation 1 (Appendix~\ref{app:profile-derivation}). The profile is permanent-heavy: 91.3\% permanent, 8.0\% intermittent, 0.75\% transient; EXPLICIT\_ERROR/PERMANENT alone carries 65.6\%, all explicit errors 71.2\%, silent content modes 19.1\%, timing modes 9.7\%. Because GitHub issues over-represent persistent, reproducible failures, we treat it as a stress profile, although its shape is in line with the 61.4\% Bohrbug share among 520 software faults from 18 JPL/NASA missions~\cite{grottke2010empirical}.

\begin{table}[t]
\small
\caption{Joint distribution of failure modes and persistence categories across 57,240 mined threads. NON\_FAILURE and UNCERTAIN persistence entries are classified but excluded from the injection profile.}
\label{tab:joint_mode_persistence}
\centering
\begin{tabular}{lccccc}
\toprule
\textbf{Failure Mode} & \textbf{Interm.} & \textbf{Perm.} & \textbf{Trans.} & \textbf{Uncert.} & \textbf{Total} \\
\midrule
DELAYED\_RESPONSE   & 51  & 157   & 1  & 24    & 233 \\
EXPLICIT\_ERROR     & 669 & 9,075 & 97 & 552   & 10,393 \\
INCORRECT\_OUTPUT   & 155 & 1,733 & 1  & 50    & 1,939 \\
INVALID\_OUTPUT     & 47  & 511   & 1  & 25    & 584 \\
MINOR\_MISTAKE      & 15  & 177   & 0  & 10    & 202 \\
SILENT\_DROP        & 167 & 963   & 4  & 83    & 1,217 \\
UNCERTAIN           & 28  & 0     & 0  & 233   & 261 \\
\midrule
\textbf{NON\_FAILURE} & 0 & 0 & 0 & 42,411 & 42,411 \\
\midrule
\textbf{Total}      & 1,132 & 12,616 & 104 & 43,388 & 57,240 \\
\bottomrule
\end{tabular}
\end{table}

\section{Experimental Protocol}
\label{sec:experiments}


\subsection{Benchmark and Task Set}
\label{sec:experiments-benchmark}

We evaluate on MCP-Bench~\cite{wang_mcp-bench_2025}, a multi-step tool-use benchmark with rubric-based LLM-as-a-judge scoring.
Using its task-generation pipeline, we produced \textbf{25 single-server tasks} from 25 successfully connected MCP servers (out of 28 attempted), each exposing multiple tools.
Tasks are presented to agents as \emph{fuzzy} (naturalistic, underspecified) user prompts; the original concrete descriptions serve only as evaluation references.
Distraction servers are disabled (\texttt{-{}-distraction-count\,0}) to isolate tool-call faults from irrelevant conversational noise.

\subsection{Experimental Factors}
\label{sec:experiments-factors}

The experiment crosses four factors in a full-factorial design.

\paragraph{LLM backbones.}
Two locally hosted open-weight models, \textbf{gpt-oss-20B} and \textbf{gpt-oss-120B}, served at their released MXFP4 precision with a fixed 16{,}000-token context window and default decoding parameters on consumer Apple Silicon via LM Studio (full hardware and serving configuration in Appendix~\ref{app:reproducibility}).
No per-model prompt tuning is applied; differences are attributable to model scale alone.

\paragraph{Agent stacks.}
Three orchestration frameworks connect to the same MCP tool interface:
(i)~a \textbf{multi-round plan-and-execute} loop from MCP-Bench~\cite{wang_mcp-bench_2025};
(ii)~an \textbf{OpenAI Agents SDK} agent~\cite{openai_agent_sdk};
(iii)~a \textbf{LangChain/LangGraph} pre-built agent~\cite{langchain_langgraph}.
Task prompts, tool schemas, context budget, and the per-task timeout/stopping policy are held constant across stacks; only intrinsic orchestration behavior (planning, retry logic, intermediate reasoning format) is allowed to vary. Appendix~\ref{app:stacks} documents these differences in detail.

\paragraph{Fault-injection conditions and repetitions.}
Each task runs under three proxy conditions: (i)~\textbf{Proxy-off}: no interposition (true baseline); (ii)~\textbf{OP0}: proxy with a null pass-through operator (exercises integration overhead); (iii)~\textbf{OPP}: proxy with the 18-rule calibrated fault profile from Section~\ref{sec:methodology} (derivation in Appendix~\ref{app:profile-derivation}). Robustness comparisons use a \emph{paired} within-task design: for every (task, seed) tuple we compute per-metric degradation $\Delta = \text{OPP} - \text{OP0}$. Three seeds per condition (fixing benchmark, injector, and framework RNG seeds) yield $2 \times 3 \times 3 \times 3 \times 25 = \mathbf{1{,}350}$ experimental cells.

\subsection{Execution and Run Policy}
\label{sec:experiments-execution}

All 1{,}350 reported cells are governed by a single run policy: a 3{,}000\,s per-task wall-clock timeout and a loop bound of 40 rounds (MCPBench \texttt{max\_execution\_rounds}), 40 turns (OpenAI \texttt{max\_turns}), or 25 supersteps (LangChain, LangGraph default; Appendix~\ref{app:stacks}), which prevents unbounded hangs under silent-drop and delay faults.
All OPP cells and every reported non-completion ran under these limits; fault-free cells that completed in earlier campaign rounds ran under tighter limits that they never reached (Appendix~\ref{app:reproducibility}), so the campaign history cannot inflate any reported non-completion.
A run terminated by the timeout, by loop-bound exhaustion, or by an unhandled tool exception is classified as a \emph{system-level failure}.
The LLM-as-a-judge evaluator (gpt-4o-mini-2024-07-18 model) scores each completed run five times with a randomized rubric order to mitigate judge instability; we report median scores.
Aggregate compute for the final 1{,}350 reported cells was approximately 58 wall-clock hours of agent execution plus about 18 hours of LLM-as-a-judge API time; the full five-round campaign history is in Appendix~\ref{app:reproducibility}.

\subsection{Metrics}
\label{sec:experiments-metrics}

We report three metric families. \textbf{(i)~Task-quality scores}: MCP-Bench's rubric-based judge rates task fulfillment, grounding, tool appropriateness, and parameter accuracy, enabling analysis beyond binary task completion. \textbf{(ii)~System-level outcomes}: timeout and unhandled-error indicators, reported as failure rates per condition. \textbf{(iii)~Latency decomposition}: end-to-end, agent-internal (LLM, planning, orchestration), and evaluation time, supporting both intrusiveness analysis under OP0 and fault-amplification analysis under OPP.

\section{Results}
\label{sec:results}

We report three primary findings: calibrated MCP faults cause severe, pervasive quality degradation (Section~\ref{sec:rq3}); framework-level error handling plausibly matters more than model scale (suggestive evidence; Section~\ref{sec:rq4}); and scale does not purchase fault tolerance (Section~\ref{sec:rq5}). Two apparatus-validation checks precondition these findings: proxy non-intrusiveness (paired Wilcoxon, all Bonferroni-corrected $p \geq 0.05$ over 18 tests; Appendix~\ref{app:rq1}) and injection fidelity (the per-run fault draw follows the profile, and cells exposed to at least one fault degrade far more than unexposed cells; Appendix~\ref{app:rq2}). Mechanistic and task-level complements appear in Appendices~\ref{app:rq3},~\ref{app:rq6}, and~\ref{app:rq7}.

\subsection{Calibrated Faults Cause Severe, Pervasive Quality Degradation}
\label{sec:rq3}

We analyze 450 OPP cells (25 tasks $\times$ 3 seeds $\times$ 3 stacks $\times$ 2 models) paired against their OP0 counterparts. Cells that timed out or errored under OPP receive 0 in the primary analysis; a completed-only secondary analysis is reported alongside.

\paragraph{Completion rates.}
Under OP0, completion rates sit between 95 and 100\% across all (stack, model) combinations (proxy-off baselines 93 to 100\%). Under OPP, they fall to 88 to 97\%, with the largest declines in OpenAI/120b (88\%) and MCPBench/120b (91\%). The 120b completes fewer tasks than the 20b in every stack (pooled 91.1\% vs.\ 95.6\%), a descriptive gap examined in Section~\ref{sec:rq5}; full per-cell rates appear in Appendix~\ref{app:rq3} (Table~\ref{tab:completion_rates}).

\paragraph{Score degradation.}
Pooled across all stacks and models, degradation is severe: task completion score drops 60\% (8.08 to 3.23), tool selection 59\%, planning effectiveness 67\%, and the remaining six metrics 54 to 68\% (full breakdown in Appendix~\ref{app:rq3}, Table~\ref{tab:itt_degradation}). Every metric drops by 4.4 to 5.2 points out of 10 (paired 95\% CIs within $\pm 0.4$ points; rank-biserial $r \approx 0.9$; Benjamini-Hochberg corrected Wilcoxon $p < 0.001$). Restricted to cells that completed under both conditions, the pattern persists (57\% drop in task completion), so degradation reflects genuine quality loss among successful runs and not only non-completion.

\begin{figure}[t]
    \centering
    \includegraphics[width=\linewidth]{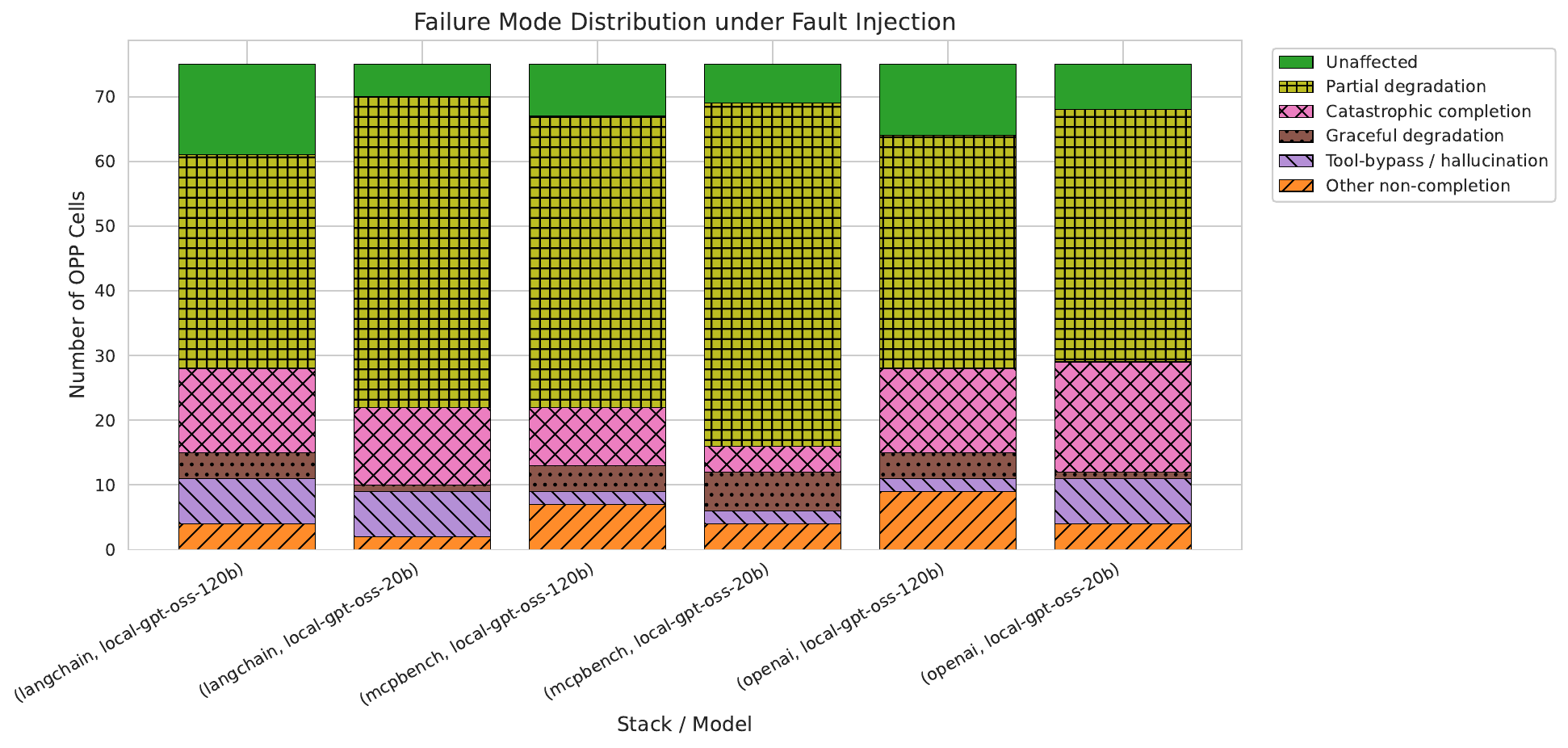}
    \caption{Failure mode distribution across all 450 OPP cells, stratified by stack and model. Partial degradation is the most common outcome (44 to 71\% of cells); non-completion, catastrophic completion, graceful degradation, and tool bypass together account for 21 to 39\%.}
    \label{fig:failure_modes}
\end{figure}

\paragraph{Failure mode taxonomy.}
Figure~\ref{fig:failure_modes} classifies all 450 OPP cells into six behavioral categories. \emph{Partial degradation} is the most common outcome (56\% of cells; 44 to 71\% across stack and model): agents typically complete tasks but with substantially reduced quality. \emph{Unaffected} cells account for 7 to 19\% per stack and model. The remaining 21 to 39\% split across catastrophic completion, graceful degradation, tool bypass, and non-completion. Annotated case studies for each category appear in Appendix~\ref{app:case-studies}.

\paragraph{Tool bypass is a baseline behavior that faults expose.}
A completed run counts as tool bypass when the judge rates task fulfillment at least 4 but tool appropriateness below 2: a plausible answer without tool evidence. Under OPP this covers 27 cells (6.0\%): 14 of 150 for LangChain, 9 for OpenAI, and 4 for MCPBench. Faults are not the main cause. The same rule flags 21 OP0 and 16 Proxy-off cells, the LangChain and MCPBench counts are identical with and without faults, and only OpenAI rises (3 to 9 cells; paired exact McNemar $p = 0.07$). Bypass also explains much of the zero-dose set: 17 of the 27 bypass cells made no tool call, so no fault could fire, and they are 17 of the 55 zero-dose OPP cells (12.2\%); excluding zero-dose cells raises the task-completion drop from 60\% to 67\%. These runs score well despite ungrounded answers. A LangChain/120b run of \texttt{unit\_converter\_000} computed every conversion from memory and scored 9.8 on task fulfillment and 9.3 on task completion, but 1.2 on tool appropriateness. Such answers inflate scores under faults while hiding what a deployment would care about, so we also report a \emph{grounded-fulfillment} view that counts task fulfillment only when tool appropriateness is at least 2. Under OPP, runs below that threshold supply 16\% of all fulfillment credit (OP0: 5\%), so grounded fulfillment falls by 63\% rather than 58\% (paired drop $-4.87$ points, 95\% CI $[-5.27, -4.47]$; OpenAI: 68\% vs.\ 61\%; Table~\ref{tab:itt_degradation}).

\subsection{Agent Stacks Shape Fault Response}
\label{sec:rq4}

\begin{figure}[t]
    \centering
    \includegraphics[width=\linewidth]{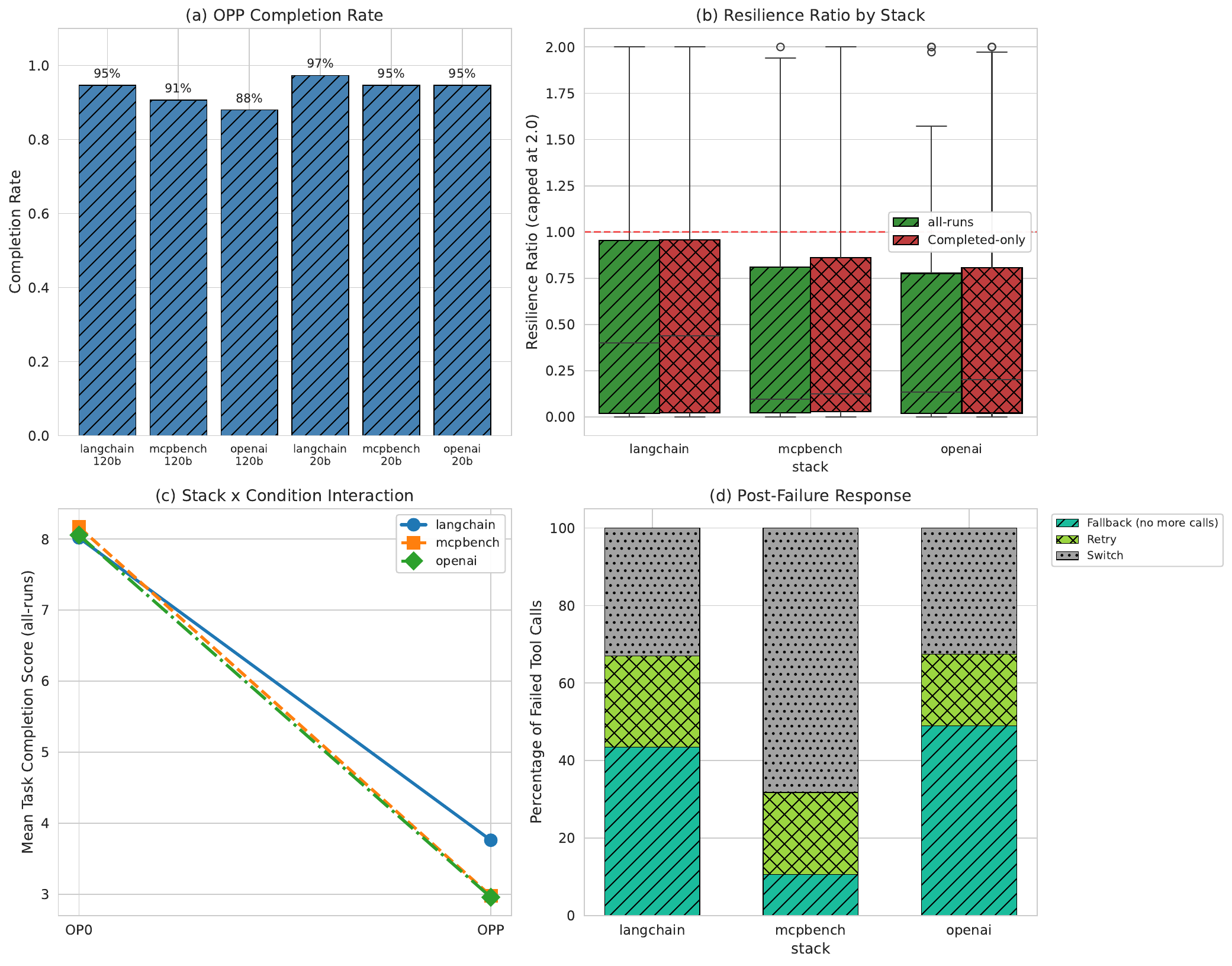}
    \caption{Stack sensitivity analysis. (a)~OPP completion rates by stack and model. (b)~Primary and completed-only resilience ratios by stack. (c)~Stack$\times$condition interaction: MCPBench and OpenAI degrade more steeply than LangChain. (d)~Post-failure response strategy differs markedly across stacks.}
    \label{fig:stack_analysis}
\end{figure}

The three agent stacks differ in how they surface tool errors to the model, producing measurably different resilience profiles.
Figure~\ref{fig:stack_analysis} summarizes the key findings.

\paragraph{Completion and resilience.}
LangChain achieves the highest OPP completion rate (95 to 97\%), OpenAI the lowest (88 to 95\%), and MCPBench is in between (91 to 95\%) (Figure~\ref{fig:stack_analysis}a). Per-task score ratios (score$_\text{OPP}$/score$_\text{OP0}$) preserve the same ordering: LangChain has the highest median, MCPBench and OpenAI are lower with wider spread, and the ordering is stable whether failed runs are scored as zero or excluded (Figure~\ref{fig:stack_analysis}b). A linear mixed-effects model with stack, condition, and their interaction as fixed effects and task as random intercept yields one significant and one marginal interaction: against a common OPP effect of $-4.26$ points, MCPBench loses a further 0.93 points relative to LangChain (95\% CI $[-1.86, -0.01]$, $p = 0.048$) and OpenAI a further 0.84 ($[-1.77, 0.08]$, $p = 0.074$) (Figure~\ref{fig:stack_analysis}c). Both intervals reach close to zero, no multiplicity correction is applied, and MCPBench also receives more fault events per cell (Appendix~\ref{app:rq2}), so we treat the framework effect as suggestive rather than established.

\paragraph{Post-failure behavior.}
The three stacks respond qualitatively differently (Figure~\ref{fig:stack_analysis}d, full analysis in Appendix~\ref{app:rq6}). MCPBench is \emph{switch-dominant}: after a failure, its planner re-plans with a different tool on the next round, consistent with its plan-and-execute loop that re-issues the full \texttt{planned\_tools} list each iteration. LangChain and OpenAI are \emph{fallback-leaning}: both exit the tool loop more often than they retry, consistent with their ReAct-style loops terminating once the LLM emits a text-only message.

\subsection{Scale Does Not Purchase Fault Tolerance}
\label{sec:rq5}

Does model scale confer fault resilience beyond raw capability?
The resilience question asks whether the 120b \emph{maintains a higher fraction of its own baseline} under OPP than the 20b does. On this benchmark, the two backbones have similar fault-free scores on completed runs (OP0: 8.37 vs.\ 8.13 task-completion score; Proxy-off: 8.11 vs.\ 8.30).

\paragraph{No detectable resilience advantage.}
In a mixed-effects model of the per-cell resilience ratio (OPP over OP0 task-completion score, non-completions scored zero; task random intercept), the model-size coefficient is indistinguishable from zero ($\beta_\text{model} = -0.009$, 95\% CI $[-0.39, 0.37]$, $p = 0.96$), and neither model$\times$stack interaction is significant ($p \geq 0.23$). The interval is wide: the data show no resilience benefit from scale for these two gpt-oss backbones under this serving setup, but cannot exclude a modest effect in either direction.

\paragraph{Decomposing the completion gap.}
Pooled OPP rates conflate three components. Fault-free quality is similar for the two backbones (above). Fault-free completion is already lower for the 120b (OP0: 96.9\% [93.7, 98.5] vs.\ 99.1\% [96.8, 99.8], Wilson 95\% CIs; Proxy-off: 96.0\% vs.\ 98.7\%). The fault-attributable drop, paired over identical (stack, task, seed) cells, is $-$5.8 points for the 120b (95\% CI $[-10.4, -1.4]$, exact McNemar $p = 0.015$) and $-$3.6 for the 20b ($[-7.2, -0.5]$, $p = 0.039$); the 2.2-point difference between them is not distinguishable from zero (95\% CI $[-3.0, 7.5]$). The pooled OPP gap, 95.6\% [92.0, 97.6] vs.\ 91.1\% [86.7, 94.2] (difference 4.4 points, 95\% CI $[-0.2, 9.3]$, Fisher exact $p = 0.088$), is therefore half baseline completion and half a larger but uncertain fault-attributable drop, and we report it as descriptive rather than as an effect of scale. The 120b is lower in every stack, most for OpenAI (88\% vs.\ 95\%) and MCPBench (91\% vs.\ 95\%), the two stacks where raw error messages reach the model most directly (Table~\ref{tab:completion_rates}). When the 120b does complete under OPP, its answers score higher (3.68 vs.\ 3.26), so capability and resilience rankings need not agree.

\paragraph{Candidate explanations.}
Greater fault exposure is consistent with the gap but does not establish it. Under OPP the 120b receives more fault injections per cell than the 20b in every stack (mean 4.6 vs.\ 3.9), it produces longer outputs under OP0 (2{,}063 vs.\ 1{,}711 output tokens per cell), and its retry loops on a failing tool are slightly longer (mean 4.5 vs.\ 3.9 consecutive retries among the 38 flagged runs; Appendix~\ref{app:rq6}). Its OPP non-completions (20 vs.\ 10) are wall-clock timeouts in all but two cases, not turn-budget exhaustion. The shared 16K-token context window and cross-stack differences in tool-call volume are co-equal alternative explanations (Section~\ref{sec:limitations}). Degradation is also strongly task-dependent: live APIs (game\_trends, $-$97\%) and deep sequential chains (national\_parks, $-$82\%) are far more vulnerable than short-chain computational tasks (unit\_converter, $-$19\%; Appendices~\ref{app:rq3},~\ref{app:rq6},~\ref{app:rq7}).

\section{Limitations}
\label{sec:limitations}

\paragraph{Construct validity.}
The LLM judge is stochastic; five-fold scoring under randomized rubric order with median aggregation reduces but does not remove its noise, and the rubric rewards some ungrounded answers (Section~\ref{sec:rq3}). The fault profile inherits the reporting bias of GitHub issues, so we treat it as a stress profile (Section~\ref{sec:methodology}). Content operators omit schema drift, payload truncation, and partial result omission.

\paragraph{Internal validity.}
The proxy is non-intrusive (Appendix~\ref{app:rq1}), and the per-run fault draw follows the profile (Appendix~\ref{app:rq2}). Two explanations remain co-equal with fault exposure: the shared 16K-token context window, within which the 120b produces longer outputs, and tool-call volume (MCPBench cells receive 8.6 fault events on average against 2.3 for LangChain and 1.8 for OpenAI; 120b cells 4.6 against 3.9). Since stacks differ in both call volume and error handling, exposure and error handling remain partially entangled, and that longer tool-call chains amplify exposure is a hypothesis, not an established mechanism (Appendix~\ref{app:rq7}). Zero-dose cells (12.2\%) and three seeds per cell limit power for per-task comparisons and rare interactions.

\paragraph{External validity.}
Both backbones come from one model family under one serving setup. MCP-Bench is easy without faults (OP0 completion 95 to 100\%, mean task-completion score 8.08, per-task means 4.8 to 9.9), which limits what the results say about harder workloads. The study is powered for the ecosystem's 91\%-permanent incident mix: TRANSIENT and INTERMITTENT entries produced 9 of 1{,}905 fired events, too few to characterize those regimes. The profile reflects the open-source MCP ecosystem as of September 2025; the stacks exclude multi-agent and human-in-the-loop workflows, and the 25 single-server tasks leave multi-server and longer-horizon settings to future benchmarks.

\section{Broader impacts}
\label{sec:broader-impacts}

\injname{} is a dual-use middleware: used as intended, it enables reproducible pre-deployment stress testing against a fault profile calibrated to the observed incident distribution of the public MCP ecosystem; as a transparent MCP interposition layer, its operator primitives (fabricated errors, delays, silent drops, perturbed payloads) could be repurposed adversarially by an attacker already positioned on the MCP path.
Since the profile is shaped by the reporting bias of public GitHub artifacts (Section~\ref{sec:methodology}), it should be treated as a stress test rather than a certification instrument.
We mitigate these risks by shipping OP0 as the default, tagging injected responses with a fault header so cooperating loggers can distinguish synthetic from real faults, and scoping documentation to pre-deployment use in staging environments.

\section{Related work}
\label{sec:related}

\paragraph{Tool-use benchmarks.}
Existing tool-use benchmarks~\cite{li_api-bank_2023, xu_tool_2023, yao2024tau, wang2024gta, mialon_gaia_2023, wang_mcp-bench_2025, chen2025acebench, liu2025mcpagentbench} measure success rates assuming tools respond correctly and on time; Xiong et al.~\cite{xiong2025butterfly} study parameter-filling failures \emph{inside} the LLM rather than the external tool, and Agent-SafetyBench~\cite{zhang2024agentsecuritybench} and Ruan et al.~\cite{ruan_identifying_2024} address adversarial threats rather than non-adversarial operational faults.
ReliabilityBench~\cite{gupta2026reliabilitybench} is closest to our work, evaluating fault tolerance under timeouts, rate limits, and schema drift; \injname{} differs in that its fault model is calibrated to the incident distribution mined from 1,774 MCP repositories, it models persistence schedules (permanent, transient, intermittent), and it operates as a transparent middleware requiring no agent modification.

\paragraph{Fault injection and chaos engineering.}
Classical fault injection evaluates system dependability by deliberately introducing faults at hardware, OS, or API boundaries~\cite{natella2016assessing,cotroneo_dependability_2021}.
Chaos engineering~\cite{7436642} extends this philosophy to production distributed systems, and Yu et al.~\cite{yu2025fit_microservices} survey its application to microservice architectures.
Recent work applies chaos engineering principles to LLM-based systems: Owotogbe~\cite{owotogbe2025chaos_llm} proposes a framework for injecting agent communication failures in multi-agent systems, and Kikuta et al.~\cite{kikuta2025chaoseater} use LLMs to automate the chaos engineering cycle on Kubernetes.
These efforts target infrastructure-level or inter-agent faults. \injname{} operates at a different layer, the MCP tool-call interface, injecting \emph{semantic} faults (content errors, timing failures, schema violations) that exercise agent-level interpretation and recovery logic.

\paragraph{MCP ecosystem reliability.}
Hou et al.~\cite{hou2025mcp_landscape} and Huang et al.~\cite{huang2026agents_fail} catalogue MCP security risks and tool-invocation failure categories post hoc, and self-correction approaches such as Reflexion~\cite{shinn2023reflexion} improve robustness but assume tools are reliable; \injname{} instead provides a controlled, reproducible mechanism to \emph{induce} realistic MCP failures and measure their impact on agent performance.

\section{Conclusion}
\label{sec:conclusion}

We presented \injname{}, a protocol-aware fault-injection middleware for MCP with a fault model calibrated to the observed incident distribution of the public MCP ecosystem (57,240 threads across 1,774 open-source repositories). Our 1,350-cell evaluation yields three findings. First, under this permanent-heavy stress profile, MCP faults cause severe quality degradation (60\% drop in task-completion score; paired difference $-4.85$ points, 95\% CI $[-5.22, -4.48]$), with \emph{partial} degradation, not catastrophic failure, the most common outcome (56\% of cells). Second, framework-level error handling appears to matter more than model scale (one significant and one marginal stack$\times$condition interaction, $p = 0.048$ and $0.074$): stacks differ in whether faults cascade or are absorbed, although this evidence is suggestive. Third, for the two gpt-oss backbones tested, scale does not purchase fault tolerance: the 120B retains no larger share of its fault-free score under faults and completes descriptively fewer tasks, with greater fault exposure a candidate explanation rather than an established mechanism. Open directions: multi-server orchestration, longer-horizon tasks, and mitigations such as fault-aware planning.

\begin{ack}
We thank SURF for providing computational resources, under allocation EINF-16024, that supported part of the experimental work in this study. The authors declare no competing interests.
\end{ack}

\bibliographystyle{plainnat}
\bibliography{chaosllm}


\appendix

\section{Fault Profile Calibration Details}
\label{app:profile-derivation}

This appendix documents how each field of the calibrated OPP profile consumed by \injname{} (file \texttt{fault\_modes/fault\_mode\_1.json}) is obtained from the mined corpus of Section~\ref{sec:methodology}. The profile contains 18 entries, one per combination of six operators (OP1 through OP6) and three injectable persistence types (INTERMITTENT, PERMANENT, TRANSIENT). Each entry specifies three kinds of values: (i)~a selection weight (\texttt{probability}), (ii)~a persistence schedule parameter (\texttt{activation\_p}, and for INTERMITTENT also \texttt{deactivation\_p}), and (iii)~operator-specific content parameters (\texttt{duration}, \texttt{delta\_ratio}, \texttt{probability\_per\_word}). We describe how each is set, separating values that are mechanically derived from the mined distribution (Table~\ref{tab:joint_mode_persistence}) from hyperparameters that are fixed on principled grounds. Table~\ref{tab:profile_derivation} lists every entry alongside its source counts.

\paragraph{Mapping failure modes to operators.}
The six operators implement one mined failure mode each: DELAYED\_RESPONSE maps to OP5, EXPLICIT\_ERROR to OP4, INCORRECT\_OUTPUT to OP2, INVALID\_OUTPUT to OP3, MINOR\_MISTAKE to OP1, and SILENT\_DROP to OP6. The UNCERTAIN mode row and the UNCERTAIN persistence column of Table~\ref{tab:joint_mode_persistence} are excluded from the profile because they carry no actionable (mode, persistence) label. The NON\_FAILURE row is also excluded, since non-failures are represented by the null injector (OP0) used in the no-fault baseline OP0.

\paragraph{Selection weight.}
For each (operator, persistence) cell, the selection weight is
\begin{equation}
\mathrm{probability} = \frac{\mathrm{count}}{N},
\end{equation}
where $N = 13{,}824$ is the sum of all injectable cells in Table~\ref{tab:joint_mode_persistence} (that is, the three injectable persistence columns for the six mined failure modes, excluding UNCERTAIN). This preserves the empirical joint distribution of (failure mode, persistence) as a probability mass function over the 18 operator-persistence entries, with the rare UNCERTAIN mass reassigned proportionally. For example, EXPLICIT\_ERROR/PERMANENT has 9{,}075 threads, yielding $9075/13824 \approx 0.6565$; MINOR\_MISTAKE/INTERMITTENT has 15 threads, yielding $15/13824 \approx 0.0011$. At injection time, \injname{} samples one entry per candidate fault episode with these weights, so the long-run frequency of every (mode, persistence) pair matches its mined prevalence.

\paragraph{INTERMITTENT activation probability.}
Equation~\eqref{eq:activation_p_derivation} was designed for a two-state schedule that alternates between faulty and correct phases with per-call transition probabilities $a = \mathtt{activation\_p}$ and $d = \mathtt{deactivation\_p}$, whose stationary faulty fraction is $a/(a+d)$; setting that fraction to the proportion of recurring (non-transient) observations of the mode classified as INTERMITTENT rather than PERMANENT gives:
\begin{equation}
\frac{a}{a+d} \;=\; \frac{\mathrm{INT}}{\mathrm{INT}+\mathrm{PERM}}
\qquad\Longrightarrow\qquad
a \;=\; d \,\cdot\, \frac{\mathrm{INT}}{\mathrm{PERM}}.
\label{eq:activation_p_derivation}
\end{equation}
We fix $d = 0.3$ across all entries (see below), which determines $a$ uniquely for each operator. For instance, DELAYED\_RESPONSE has $\mathrm{INT}/\mathrm{PERM} = 51/157$, giving $a = 0.3 \cdot 51/157 \approx 0.0975$. All six INTERMITTENT activation probabilities in the profile are reproduced by Equation~\eqref{eq:activation_p_derivation} to four decimal places. The shipped injector, however, uses these parameters differently: once an INTERMITTENT entry is drawn, each call fires independently with probability $a$, and from the next call on the entry is released with probability $d$ per call, after which a new entry is drawn from the profile. The faulty fraction while the entry is held is therefore $a$ rather than $a/(a+d)$, and an episode spans $1 + 1/d \approx 4.3$ calls on average, so it is expected to fire $a(1 + 1/d) \le 0.42$ times. The six activation probabilities are data-derived in the sense of Equation~\eqref{eq:activation_p_derivation}, but under the implemented schedule INTERMITTENT entries fire far less often than the mined INT/(INT+PERM) ratios imply, consistent with the 6 INTERMITTENT events in the fired-faults log (Appendix~\ref{app:rq2}).

\paragraph{TRANSIENT activation probability.}
TRANSIENT faults, by definition, fire on a single tool call and then reset (Section~\ref{sec:design-fault-api}). Once a TRANSIENT entry is selected, the corresponding call is faulty with certainty, so $\mathtt{activation\_p} = 1.0$ for every TRANSIENT entry. The mined frequency of each (mode, TRANSIENT) cell is reflected in the selection weight, not in the schedule parameter.

\paragraph{PERMANENT schedule.}
A PERMANENT entry has no activation parameter: once the entry is selected, the operator fires on every subsequent matching call (Section~\ref{sec:design-fault-api}).

\paragraph{Choice of $d=0.3$ for INTERMITTENT deactivation.}
Equation~\eqref{eq:activation_p_derivation} identifies only the ratio $a/d$ from data. The mined threads do not provide reliable episode-length statistics (GitHub issues rarely report call-level burst lengths), so $d$ is a free hyperparameter. In the two-state design it sets the mean faulty burst to $1/d \approx 3.3$ calls; in the shipped injector it sets how long a drawn INTERMITTENT entry is held ($1 + 1/d \approx 4.3$ calls on average). Either way the episode is short relative to the execution loops (up to 25 or 40 rounds per task), so a single INTERMITTENT episode does not monopolize a task execution. Choosing a common $d$ for all operators keeps the calibration one-dimensional per mode and ensures that cross-operator comparisons of INTERMITTENT effects are not confounded by different episode lengths. The loop bounds themselves differ across stacks and are documented in Appendix~\ref{app:stacks}.

\paragraph{OP5 delay duration.}
The MCP HTTP transport timeout is 60 seconds (Section~\ref{sec:experiments-benchmark}, default in \texttt{benchmark\_config.yaml}). We set $\mathtt{duration} = 30$ seconds: half the HTTP timeout, which is an order of magnitude above typical healthy tool latency (a few seconds) and therefore clearly disruptive, but strictly below the transport cutoff so that OP5 remains distinct from OP6 (SILENT\_DROP) and from OP4 (EXPLICIT\_ERROR) at the protocol boundary. Longer values would cause transport-level timeouts and collapse DELAYED\_RESPONSE into SILENT\_DROP; shorter values would risk being absorbed by the agent's natural per-call variance and not register as a delay.

\paragraph{OP1 and OP2 content-perturbation parameters.}
OP1 (MINOR\_MISTAKE) and OP2 (INCORRECT\_OUTPUT) both act word-by-word on the response, but realize qualitatively different failure modes. OP1 mutates each selected word by a single-character substitution of magnitude controlled by \texttt{delta\_ratio}; OP2 replaces each selected word by a random value of the same type. The parameters are chosen to reflect the semantic separation between the two modes.
OP1 uses $\mathtt{probability\_per\_word} = 0.08$ and $\mathtt{delta\_ratio} = 0.05$, yielding sparse, low-magnitude perturbations that leave most of the response intact, consistent with the MINOR\_MISTAKE label (small deviation, structurally valid). The 5\% delta level is the canonical minor-perturbation intensity used in perturbation-robustness literature.
OP2 uses $\mathtt{probability\_per\_word} = 0.35$, which corrupts roughly one third of the response, consistent with INCORRECT\_OUTPUT (semantically wrong but structurally valid). We avoid values close to 1 because a fully garbled response typically fails downstream JSON parsing, collapsing OP2 into OP3 (INVALID\_OUTPUT). The 0.08 vs.\ 0.35 gap preserves a clear intensity ordering between the two content operators; the mined corpus does not support finer tuning because threads rarely quantify the numeric extent of a "minor" versus "wrong" response.

\paragraph{OP3, OP4, OP6 parameters.}
These operators take no tunable parameters. OP3 deterministically casts the response to a type that violates the tool's schema, OP4 raises a structured MCP error, and OP6 suspends the call indefinitely. Their behavior is fully specified by the persistence schedule and requires no additional calibration.

\paragraph{Summary.}
Equation~\eqref{eq:activation_p_derivation} and $\mathrm{probability} = \mathrm{count}/N$ are the two deterministic rules that map the mined distribution onto the JSON profile. The remaining values ($d=0.3$; $\mathtt{duration}=30$; $\mathtt{delta\_ratio}=0.05$; $\mathtt{probability\_per\_word} \in \{0.08, 0.35\}$) are fixed hyperparameters whose roles are to (i)~set the temporal scale of INTERMITTENT episodes, and (ii)~separate content operators at semantically meaningful intensities without collapsing them into adjacent modes. Together these choices produce a fully specified, seedable fault profile whose selection weights are traceable to Table~\ref{tab:joint_mode_persistence}.

\begin{table}[h]
\centering
\small
\caption{Derivation of every entry in the OPP profile (\texttt{fault\_mode\_1.json}). Selection weights are $\mathrm{count}/N$ with $N=13{,}824$. INTERMITTENT activation probabilities follow Equation~\eqref{eq:activation_p_derivation} with $d=0.3$; TRANSIENT activation is fixed at $1.0$. Small residuals reflect four-decimal rounding in the shipped JSON.}
\label{tab:profile_derivation}
\begin{tabular}{@{}llrrrl@{}}
\toprule
\textbf{Operator (mode)} & \textbf{Persistence} & \textbf{Count} & \textbf{prob.} & \textbf{act.\,p} & \textbf{Derivation of act.\,p} \\
\midrule
OP5 (DELAYED\_RESPONSE)    & INTERMITTENT &    51 & 0.0037 & 0.0975 & $0.3 \cdot 51/157$  \\
                           & PERMANENT    &   157 & 0.0114 & n/a    & n/a                 \\
                           & TRANSIENT    &     1 & 0.0001 & 1.0    & fixed               \\
\midrule
OP4 (EXPLICIT\_ERROR)      & INTERMITTENT &   669 & 0.0484 & 0.0221 & $0.3 \cdot 669/9075$ \\
                           & PERMANENT    &  9075 & 0.6565 & n/a    & n/a                 \\
                           & TRANSIENT    &    97 & 0.0070 & 1.0    & fixed               \\
\midrule
OP2 (INCORRECT\_OUTPUT)    & INTERMITTENT &   155 & 0.0112 & 0.0268 & $0.3 \cdot 155/1733$ \\
                           & PERMANENT    &  1733 & 0.1254 & n/a    & n/a                 \\
                           & TRANSIENT    &     1 & 0.0001 & 1.0    & fixed               \\
\midrule
OP3 (INVALID\_OUTPUT)      & INTERMITTENT &    47 & 0.0034 & 0.0276 & $0.3 \cdot 47/511$  \\
                           & PERMANENT    &   511 & 0.0370 & n/a    & n/a                 \\
                           & TRANSIENT    &     1 & 0.0001 & 1.0    & fixed               \\
\midrule
OP1 (MINOR\_MISTAKE)       & INTERMITTENT &    15 & 0.0011 & 0.0254 & $0.3 \cdot 15/177$  \\
                           & PERMANENT    &   177 & 0.0128 & n/a    & n/a                 \\
                           & TRANSIENT    &     0 & 0.0000 & 1.0    & fixed (unused)      \\
\midrule
OP6 (SILENT\_DROP)         & INTERMITTENT &   167 & 0.0121 & 0.0520 & $0.3 \cdot 167/963$ \\
                           & PERMANENT    &   963 & 0.0697 & n/a    & n/a                 \\
                           & TRANSIENT    &     4 & 0.0003 & 1.0    & fixed               \\
\bottomrule
\end{tabular}
\end{table}

\section{Proxy Non-Intrusiveness: Full Analysis}
\label{app:rq1}

The experimental design requires that the proxy with null injection (OP0) is outcome-equivalent to the no-proxy baseline (proxy-off).
We test this by comparing task-quality scores and execution latency across all six (stack, model) combinations.

\paragraph{Score equivalence.}
For each (stack, model) pair, we run paired Wilcoxon signed-rank tests on three composite metrics (task completion, tool selection, planning effectiveness) between proxy-off and OP0, pairing by (task, seed).
After Bonferroni correction across 18 tests (6 pairs $\times$ 3 metrics), no comparison reaches significance at $\alpha = 0.05$.
The smallest uncorrected $p$ is mcpbench/20b on planning effectiveness ($p_\text{uncorr} = 0.003$, $r = 0.40$, Bonferroni $p = 0.050$), followed by mcpbench/20b on tool selection ($p_\text{uncorr} = 0.006$, $r = 0.39$, Bonferroni $p = 0.10$); all three mcpbench/20b deltas are small and negative ($-0.27$ to $-0.45$ points), so a slight proxy effect in that cell cannot be excluded.
Mean score deltas are small: the largest absolute difference is 0.55 points (openai/120b task completion), with no consistent direction across stacks, confirming that the proxy does not systematically bias scores.

\begin{figure}[h]
    \centering
    \includegraphics[width=0.85\linewidth]{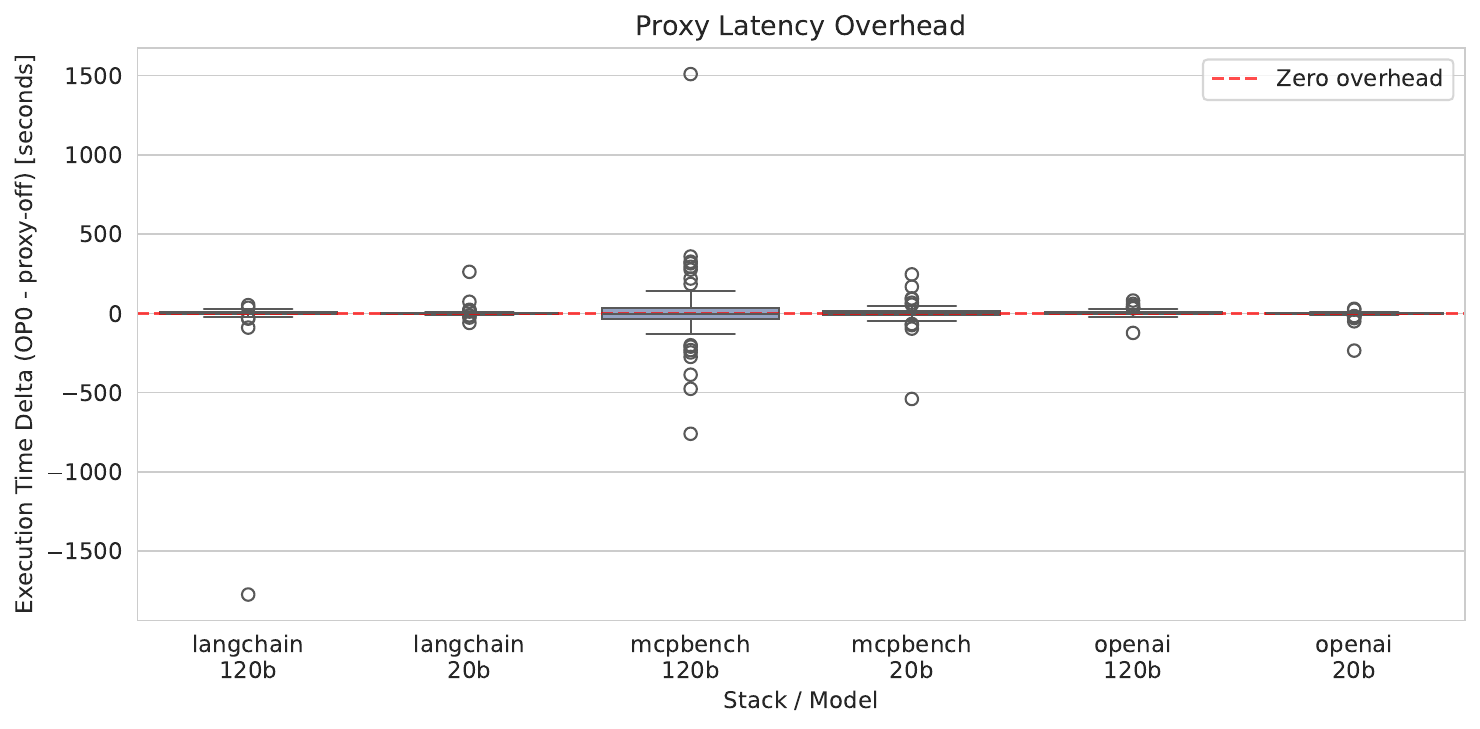}
    \caption{Distribution of execution-time deltas (OP0 $-$ proxy-off) per task, by stack. Distributions are centered near zero with no systematic positive bias, confirming minimal proxy overhead.}
    \label{fig:rq1_latency}
\end{figure}

\paragraph{Latency overhead.}
Mean execution-time deltas (OP0 minus proxy-off) range from $-$22\,s (langchain/120b) to $+$7\,s (mcpbench/120b) in absolute seconds, and from $-$12\% (openai/20b) to $+$21\% (langchain/20b) as a fraction of the proxy-off baseline.
The negative delta for LangChain reflects run-to-run variance in LLM inference time rather than proxy-induced speedup.
Expressed as per-tool-call overhead, the proxy adds an estimated 0.1 to 0.5\,s per tool invocation, which is negligible relative to LLM inference latency (typically 2 to 15\,s per turn).
Figure~\ref{fig:rq1_latency} shows the distribution of latency deltas by stack.

\paragraph{Conclusion.}
The proxy satisfies the non-intrusiveness requirement: OP0 is a valid fault-free baseline, and all subsequent OP0-vs-OPP comparisons are not confounded by instrumentation artifacts.

\section{Injection Fidelity: Full Analysis}
\label{app:rq2}

This section checks how the specified profile translates into realized fault events under OPP and whether fault exposure is sufficient to support causal claims.
We define the \emph{fault dose} of an experimental cell as the number of fault injection events fired by the proxy during that cell's execution, computed by counting the cell's entries in the fired-faults log. Dose ranges from zero (when the agent makes no tool call, or draws an INTERMITTENT entry that never fires) to several dozen, depending on tool-call volume and the persistence state of each triggered fault.

\begin{figure}[h]
    \centering
    \includegraphics[width=0.85\linewidth]{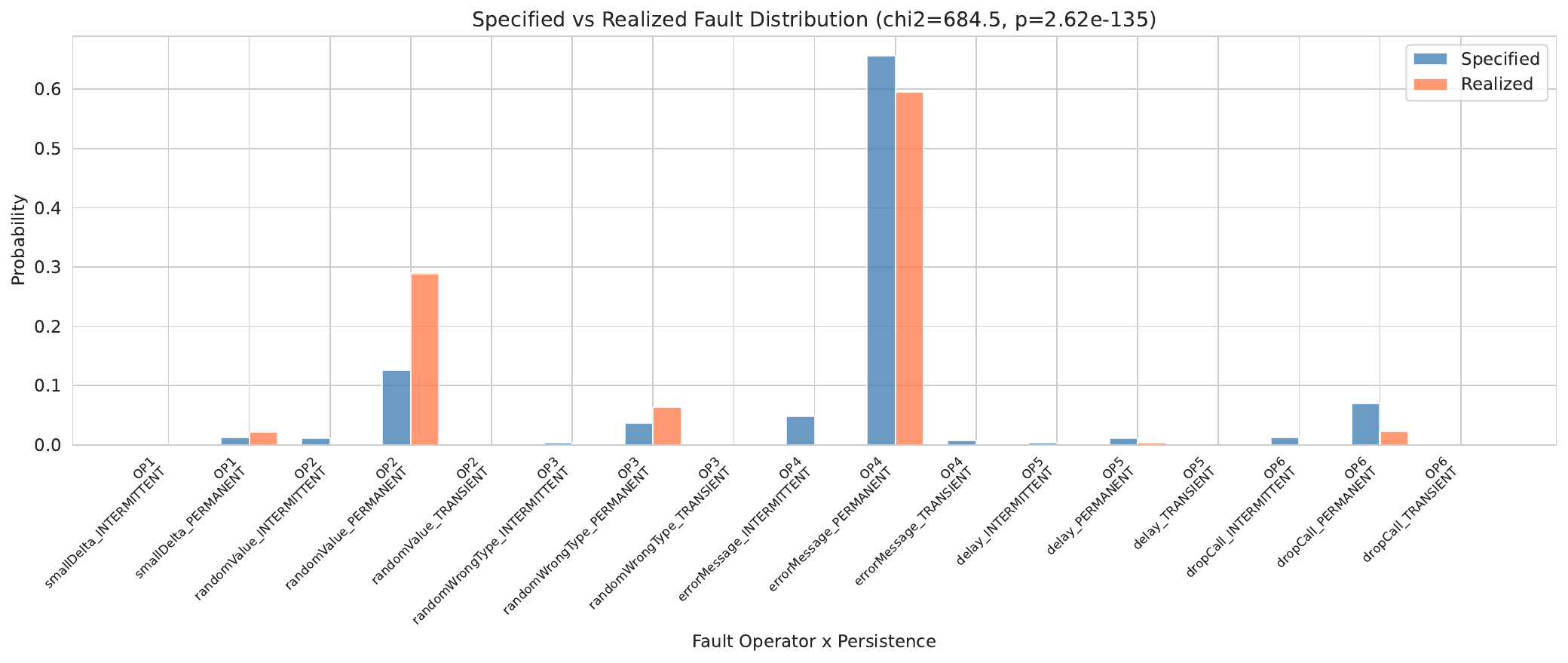}
    \caption{Specified selection weights vs.\ realized share of fired events (1{,}905 events from 395 OPP cells) across (operator $\times$ persistence) entries. Because a PERMANENT draw fires on every subsequent tool call, entries that sustain longer tool-call sequences (notably OP2/PERMANENT) are over-represented per event, and INTERMITTENT entries rarely fire. The title reports the per-event goodness-of-fit test.}
    \label{fig:rq2_fidelity}
\end{figure}

\paragraph{Distribution fidelity.}
We compare the specified (fault operator $\times$ persistence) selection weights with the realized frequencies in the fired-faults log: 1{,}905 fired events from the 395 OPP cells with at least one injection (Figure~\ref{fig:rq2_fidelity}).
The per-event realized mix departs detectably from the specified weights ($\chi^2 = 684.5$, $\mathrm{df} = 16$ over the 17 entries with non-zero weight, $p < 0.001$; Cram\'er's $V = 0.15$; total variation distance $0.20$).
EXPLICIT\_ERROR/PERMANENT remains the dominant entry (65.6\% specified, 59.5\% realized), but INCORRECT\_OUTPUT/PERMANENT (OP2) rises from 12.5\% to 28.9\%; PERMANENT entries account for 99.5\% of fired events against 91.3\% of the weight, and INTERMITTENT entries (8.0\% of the weight) fire only 6 times (0.3\%).
The departure reflects how the profile is consumed rather than a sampling error.
Selection weights apply per draw, not per event: the injector draws an entry at the first tool call of a run, and a PERMANENT draw then fires on every subsequent call, so each such run contributes as many events as it makes tool calls.
Event volume per draw depends on the entry and the stack: runs under OP2/PERMANENT, whose perturbed outputs return as successes, average 7.5 fired events, against 4.2 under EXPLICIT\_ERROR/PERMANENT, a gap that holds within every stack; and MCPBench runs contribute 68\% of all events (mean 8.6 per cell, versus 2.3 for LangChain and 1.8 for OpenAI).
INTERMITTENT entries fire on a given call only with activation probability $a \le 0.10$, and the median OPP cell makes two tool calls, so these entries rarely fire.
At the level of the draw itself, the first fired entry of each run is consistent with the specified weights renormalized over the PERMANENT and TRANSIENT entries (391 runs; Monte Carlo $p = 0.22$, $V = 0.06$), although OP2/PERMANENT is drawn more often than expected (74 runs against 53).
The profile thus fixes what is drawn, while persistence and agent behavior determine how many faults each draw delivers.
Analyses that condition on exposure (zero-dose cells and exposure versus dose, below) use realized doses from the fired-faults log, and the OPP versus OP0 contrasts of Section~\ref{sec:rq3} compare conditions as assigned, so neither depends on the per-event mix matching the profile.

\paragraph{Zero-dose runs.}
Of 450 OPP cells, 55 (12.2\%) received no fault. Every profile entry targets all tools and every PERMANENT or TRANSIENT draw fires on its first call, so a cell is zero-dose only if the agent made no tool call (29 cells, 17 of them classified as tool bypass) or drew an INTERMITTENT entry that never fired (26 cells, consistent with the 8\% INTERMITTENT selection weight; Appendix~\ref{app:profile-derivation}).
These cells are distributed across all stacks (LangChain: 21, MCPBench: 15, OpenAI: 19) and are concentrated in tasks where alternative solution paths exist (unit\_converter, math\_mcp, wikipedia).
Including zero-dose cells in the primary analysis dilutes the measured fault effect; excluding them amplifies the degradation from $-$60\% to $-$67\% in the task completion score.

\paragraph{Exposure and dose.}
Exposure to at least one fault is strongly associated with degradation: OPP cells with one or more fired faults score 2.72 on task completion (OP0: 8.15 for the same cells), against 6.93 for zero-dose cells (OP0: 7.61). The \emph{number} of faults is not. In a linear mixed-effects model \texttt{score $\sim$ fault\_dose + stack + (1\,|\,task\_id)} fit by REML to the 366 completed OPP cells with nonzero dose, the dose coefficient is small and, if anything, positive for all nine metrics (task completion: $+0.06$ points per fired fault, 95\% CI $[-0.004, 0.12]$; including zero-dose cells: $+0.01$, $[-0.05, 0.07]$). Two features of the design explain this. Most runs receive a single PERMANENT draw that fires from the first tool call (see distribution fidelity above), so degradation sets in at first exposure; and dose is confounded with tool-call volume, since agents that keep calling tools, for instance under silent content faults, accumulate more events while still producing partial answers. Zero-dose cells are not a randomized control either: 29 of the 55 made no tool call, and their fault-free scores are lower. We therefore read the evidence as an exposure effect rather than a dose-response relationship.

\section{Supplementary Analyses of Fault Impact}
\label{app:rq3}

\paragraph{Per-cell completion rates.}
Table~\ref{tab:completion_rates} reports task completion rates for every (model, stack, condition) cell, expanding the headline numbers in Section~\ref{sec:rq3}. OP0 rates are 95 to 100\% and proxy-off rates 93 to 100\% across all (stack, model) cells; the largest OPP drops are for the 120b under MCPBench and OpenAI, although per-stack intervals overlap widely.

\begin{table}[h]
\centering
\caption{Task completion rates (\%) by model, stack, and proxy condition, with Wilson 95\% CIs ($n = 75$ cells per stack row, 225 per pooled row). Per-stack intervals overlap widely; the largest OPP drops are for the 120b under MCPBench and OpenAI.}
\label{tab:completion_rates}
\small
\begin{tabular}{@{}llccc@{}}
\toprule
\textbf{Model} & \textbf{Stack} & \textbf{Proxy-off} & \textbf{OP0} & \textbf{OPP} \\
\midrule
\multirow{4}{*}{20b}   & LangChain & 100.0 [95.1, 100.0] & 100.0 [95.1, 100.0] & 97.3 [90.8, 99.3] \\
                       & MCPBench  & 100.0 [95.1, 100.0] & 100.0 [95.1, 100.0] & 94.7 [87.1, 97.9] \\
                       & OpenAI    & 96.0 [88.9, 98.6] & 97.3 [90.8, 99.3] & 94.7 [87.1, 97.9] \\
\cmidrule(l){2-5}
                       & Pooled    & 98.7 [96.2, 99.5] & 99.1 [96.8, 99.8] & 95.6 [92.0, 97.6] \\
\midrule
\multirow{4}{*}{120b}  & LangChain & 94.7 [87.1, 97.9] & 96.0 [88.9, 98.6] & 94.7 [87.1, 97.9] \\
                       & MCPBench  & 100.0 [95.1, 100.0] & 100.0 [95.1, 100.0] & 90.7 [82.0, 95.4] \\
                       & OpenAI    & 93.3 [85.3, 97.1] & 94.7 [87.1, 97.9] & 88.0 [78.7, 93.6] \\
\cmidrule(l){2-5}
                       & Pooled    & 96.0 [92.6, 97.9] & 96.9 [93.7, 98.5] & 91.1 [86.7, 94.2] \\
\bottomrule
\end{tabular}
\end{table}

\paragraph{Per-metric score degradation.}
Table~\ref{tab:itt_degradation} reports the full nine-metric degradation summary referenced in Section~\ref{sec:rq3}.

\begin{table}[h]
\centering
\caption{Score degradation under fault injection (OPP vs.\ OP0), pooled across stacks and models. Paired drops have 95\% CIs within $\pm 0.4$ points and Benjamini-Hochberg corrected Wilcoxon $p < 0.001$ for every metric. $^{\dagger}$Task fulfillment counted only when tool appropriateness is at least 2 (Section~\ref{sec:rq3}).}
\label{tab:itt_degradation}
\small
\begin{tabular}{@{}lcccc@{}}
\toprule
\textbf{Metric} & \textbf{OP0} & \textbf{OPP} & \textbf{$\Delta$} & \textbf{\% Drop} \\
\midrule
Task completion     & 8.08 & 3.23 & $-$4.85 & $-$60 \\
Tool selection      & 8.10 & 3.36 & $-$4.74 & $-$59 \\
Planning eff.       & 7.51 & 2.49 & $-$5.02 & $-$67 \\
Task fulfillment    & 8.07 & 3.37 & $-$4.70 & $-$58 \\
Grounded fulfillment$^{\dagger}$ & 7.70 & 2.83 & $-$4.87 & $-$63 \\
Grounding           & 8.09 & 3.10 & $-$5.00 & $-$62 \\
Tool appropriateness & 8.21 & 3.81 & $-$4.40 & $-$54 \\
Parameter accuracy  & 8.00 & 2.91 & $-$5.08 & $-$64 \\
Dependency awareness & 7.94 & 2.75 & $-$5.19 & $-$65 \\
Parallelism eff.    & 7.08 & 2.23 & $-$4.85 & $-$68 \\
\bottomrule
\end{tabular}
\end{table}

\begin{figure}[h]
    \centering
    \includegraphics[width=0.75\linewidth]{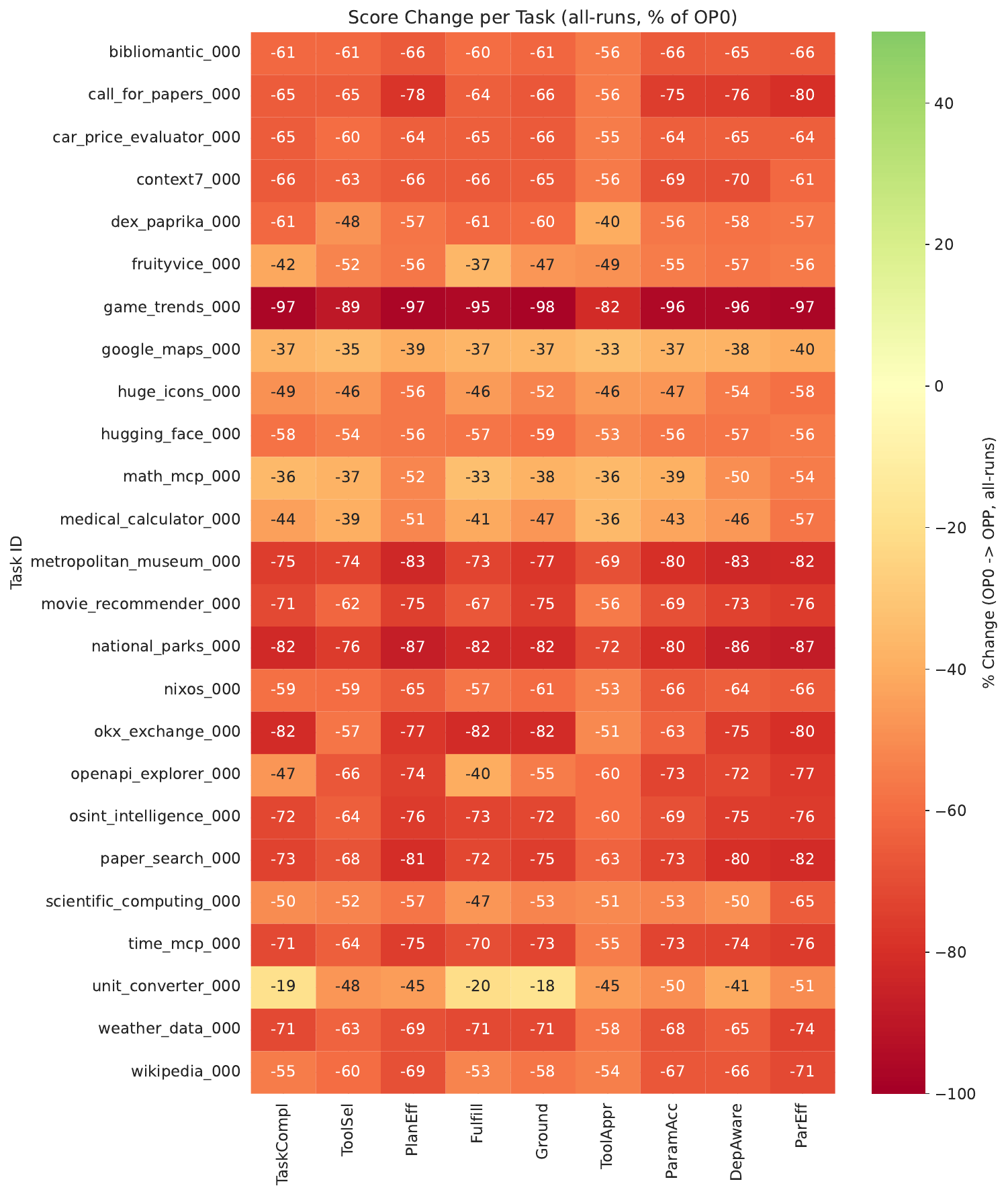}
    \caption{Per-task score degradation (OPP vs.\ OP0, \% change) across all nine evaluation metrics. Tasks are ordered alphabetically. Darker cells indicate more severe degradation.}
    \label{fig:score_heatmap}
\end{figure}

\paragraph{Per-task score heatmap.}
Figure~\ref{fig:score_heatmap} shows percentage degradation for each of the 25 tasks across all nine evaluation metrics, pooled across stacks, models, and seeds.
Degradation ranges from $-$19\% (unit\_converter, task completion) to $-$98\% (game\_trends, grounding).
Tasks involving live APIs or deep sequential chains are consistently the most vulnerable, while short-chain computational tasks show the most resilience.

\paragraph{Completed-only analysis.}
Restricting to cells that completed under both OP0 and OPP, task completion score drops by 57\% (from 8.26 to 3.53), compared to 60\% in the primary analysis.
The gap between all-in and completed-only is modest (3 percentage points), indicating that non-completion, while present, is not the dominant driver of score degradation; quality loss among completed runs accounts for most of the effect.

\paragraph{Case studies.}
We present detailed case studies for all six failure categories in Appendix~\ref{app:case-studies}, with representative reasoning traces illustrating how agents cope with injected faults.

\section{Failure Mode Case Studies}
\label{app:case-studies}

To ground the failure mode taxonomy (Section~\ref{sec:rq3}), we present annotated case studies for each of the six categories.
All examples are drawn from the langchain/120b condition under OPP (seed~1), which exhibits the widest behavioral diversity.
Scores are reported on the 0 to 10 scale; ``nan'' indicates system failure (non-completion).

\paragraph{Non-completion.}
\texttt{car\_price\_evaluator\_000} and \texttt{metropolitan\_museum\_000} both failed outright, producing no scorable output (all metrics nan).
In both cases, the agent encountered persistent access-denied errors and exhausted its turn budget retrying.
The reasoning trace for \texttt{car\_price\_evaluator\_000} reveals repeated attempts with minor parameter variations: ``\emph{Since previous attempts denied access, maybe still can try? We have only three tools all denied previously. Might be permission issue; perhaps we cannot get data. Could stop.}''
The agent recognized the futility of retrying but continued anyway, a pattern consistent with the absence of explicit fault-type diagnosis noted in Appendix~\ref{app:rq6}.

\paragraph{Tool-bypass / hallucination.}
\texttt{openapi\_explorer\_000} completed with a task score of 4.4 but tool appropriateness of only 0.8: the agent produced a detailed comparison of the GitHub and GitLab REST APIs entirely from parametric knowledge, never successfully calling the designated OpenAPI exploration tool.
More strikingly, \texttt{unit\_converter\_000} scored 9.8 on task fulfillment (9.3 task completion) by computing unit conversions from memory without any tool call, receiving only 1.2 on tool appropriateness; since it made no call, no fault reached it.
These cases are qualitatively concerning because a downstream consumer would see plausible, high-quality output with no indication that the designated tools were never consulted.

\paragraph{Graceful degradation.}
\texttt{google\_maps\_000} exemplifies this category: the agent acknowledged tool unavailability and offered alternative approaches (``\emph{It looks like I'm currently unable to query the nearby-search tool. If you'd like, I can try a different approach}''), but received 0.0 across all metrics because it produced no tool-grounded answer.
\texttt{context7\_000} similarly degraded gracefully, falling back to parametric knowledge about Express middleware with a task score of 0.2.
The reasoning trace shows explicit recognition of the problem: ``\emph{Since resolve-library-id failed due access denied, maybe cannot use tools? We might skip and answer from knowledge.}''
These cases highlight a tension in current evaluation rubrics: scores penalize tool non-use even when the agent's behavior is arguably adaptive.

\paragraph{Catastrophic completion.}
\texttt{bibliomantic\_000} and \texttt{metropolitan\_museum\_000} (seed~1) both completed but scored 0.0 across all metrics.
The agent produced polite refusals (``\emph{I'm sorry, but I don't have the ability to run the I~Ching-related tools}'') that technically constitute a response but provide no task-relevant content.
The reasoning traces for \texttt{bibliomantic\_000} are particularly revealing, showing four consecutive planning iterations where the agent hypothesizes parameter format issues: ``\emph{The previous attempts failed due to output format issues; maybe we need proper parameters.  For i\_ching\_divination tool expects query optional string.}''
At no point does the agent recognize that the fault is permanent rather than a transient parameter mismatch.

\paragraph{Partial degradation.}
\texttt{dex\_paprika\_000} (task score 0.4, tool appropriateness 2.2) and \texttt{fruityvice\_000} (task score 1.9, tool appropriateness 4.0) represent the modal outcome under fault injection.
In both cases, the agent made some successful tool calls but produced incomplete or partially incorrect answers.
\texttt{fruityvice\_000} illustrates a common pattern: the agent retrieved partial data about fruit nutritional content, supplemented it with parametric knowledge of variable accuracy, and produced a mixed-quality response that scores well on structure but poorly on grounding.

\paragraph{Unaffected.}
\texttt{call\_for\_papers\_000} (task score 9.4, tool appropriateness 9.8) and \texttt{huge\_icons\_000} (task score 10.0, tool appropriateness 10.0) completed with near-perfect scores despite running under OPP.
For \texttt{call\_for\_papers\_000}, the reasoning trace shows the agent quickly pivoting after tool failure: ``\emph{Since get\_events is denied maybe we can't use? \ldots\ User wants info on upcoming AI conferences \ldots\ Since tool calls failed, I must rely on my knowledge.}''
Despite falling back to parametric knowledge, the agent produced a comprehensive, accurate answer because the task domain (well-known conference information) is well-represented in training data.
\texttt{huge\_icons\_000} succeeded because the injected faults did not affect the specific tool calls required for the task (the fault was semantically benign for that call pattern).
These cases underscore that resilience is task-dependent: the same fault profile can be devastating for one task and irrelevant for another.

\paragraph{Cross-cutting observations.}
Three patterns recur across categories.
First, agents universally treat failures as transient, retrying with minor parameter variations rather than diagnosing fault persistence.
Second, the quality of fallback behavior depends heavily on whether the task domain overlaps with the model's training data: API-dependent tasks with specialized schemas degrade catastrophically, while general-knowledge tasks degrade gracefully or not at all.
Third, current evaluation rubrics do not distinguish between ``correctly identified tool unavailability and provided best-effort answer'' and ``failed to use tools,'' conflating adaptive and maladaptive behaviors under the same low score.

\section{Agent Stack Orchestration Details}
\label{app:stacks}

This appendix documents the intrinsic orchestration behaviors that differ across the three agent stacks and that we allow to vary by design (Section~\ref{sec:experiments-factors}).
Task prompts, tool schemas, LLM backbones, decoding parameters, context budget, per-task wall-clock timeout, and the fault-injection layer are held constant.
The differences below are relevant for interpreting the post-failure behavior in Section~\ref{sec:rq4} and the retry-loop analysis in Appendix~\ref{app:rq6}.
We retain the labels \textbf{MCPBench}, \textbf{OpenAI}, \textbf{LangChain} from the main text; these follow the naming of the reference implementations rather than any claim about the orchestration pattern per se.

\paragraph{MCPBench executor.}
The reference MCP-Bench agent~\cite{wang_mcp-bench_2025} is a multi-round \emph{plan-and-execute} loop rather than a per-step reason-act-observe loop.
On every round, a planner LLM call is asked to emit a strict JSON object with three keys: a free-text \texttt{reasoning} field, a Boolean \texttt{should\_continue} gate, and a list \texttt{planned\_tools} of independent tool calls.
The executor then dispatches the planned calls concurrently with a per-server semaphore.
Tool-call errors are surfaced back to the next planner call as part of an \emph{accumulated information} buffer, which is LLM-summarized when it approaches the context budget.
Retries exist at the LLM-call level (task/round nesting, token-budget fallbacks, format-fix re-prompts) but not at the tool-call level; whether to retry or switch tools is the planner LLM's decision.

\paragraph{OpenAI (Agents SDK).}
We use the OpenAI Agents SDK's \texttt{Runner.run} with \texttt{tool\_choice="required"} on the chat-completions API, which gives the model native OpenAI function calling and parallel tool calls by default.
The SDK runs a single message loop: each turn produces an assistant message optionally containing one or more \texttt{tool\_calls}, the SDK executes them and appends \texttt{tool}-role messages with the results, and the loop terminates when the assistant message contains only text.
MCP tools are exposed through a thin adapter that maps tool errors to a \texttt{CallToolResult} with \texttt{isError=True}, so the LLM sees a structured error message on the next turn and can retry, switch, or stop.
No custom retry logic is added on top of the SDK.

\paragraph{LangChain (\texttt{create\_agent}).}
The LangChain stack uses the v1 \texttt{create\_agent} abstraction, which compiles a ReAct state machine onto LangGraph: alternating \texttt{agent} (LLM) and \texttt{tools} nodes until the LLM stops emitting \texttt{tool\_calls}~\cite{langchain_langgraph}.
Multiple tool calls in a single assistant message are executed in parallel by \texttt{ToolNode} (internally, \texttt{asyncio.gather}).
A lightweight tool-monitoring middleware wraps each tool invocation so that exceptions are caught and returned as a \texttt{ToolMessage} payload, analogous to the \texttt{isError} path in the OpenAI adapter.
As in the other stacks, any retry or tool-switching is decided by the LLM on the next agent step.

\paragraph{Loop bounds.}
For the MCPBench and OpenAI stacks, we cap the orchestration loop at 40 iterations (\texttt{max\_execution\_rounds=40} and \texttt{max\_turns=40} respectively).
The LangChain stack inherits LangGraph's default \texttt{recursion\_limit} of 25 supersteps, which corresponds to roughly 12 agent/tool cycles.
This asymmetry is a property of the stack's defaults rather than a controlled variable, and it is conservative for our analysis: a tighter loop could only \emph{decrease} LangChain's observed retry and tool-switching rates, so it cannot explain the fact that LangChain exhibits the \emph{highest} OPP completion rate in Section~\ref{sec:rq4}.
We also cap each run at a shared 3000-second wall-clock timeout. The timeout binds before either MCPBench or LangChain saturates its loop bound; a small fraction of OpenAI runs (14 of 450, about 3\%) do reach \texttt{max\_turns=40} before the timeout fires.

\paragraph{Implications for stack sensitivity and behavioral adaptation.}
Table~\ref{tab:stack_details} summarizes the intrinsic orchestration differences across the three stacks.
The stacks differ most in how the decision to stop further tool calls is represented, and this shapes the post-failure profile reported in Section~\ref{sec:rq4}.
MCPBench drives the loop through a per-round planner that re-plans the full tool list on every round, so after a fault the natural continuation is to plan a different tool; the result is a switch-dominant profile (68\% switch, 21\% retry, 11\% fallback).
OpenAI and LangChain both encode ``stop'' implicitly as ``emit an assistant message with no tool calls'', so once the LLM runs out of hypotheses about the failing tool the path of least resistance is to verbalize a final answer; both stacks are fallback-leaning (49\% and 44\% fallback, respectively).
LangChain occupies the middle ground on tool-switching (33\%) and shows the highest retry rate (24\%), consistent with its ReAct loop re-entering the agent node after each tool error.
The richer LLM-call retry machinery in the MCPBench stack, combined with its higher per-run tool-call count (mean 8.3 under OPP versus 2.5 for LangChain and 1.9 for OpenAI; Appendix~\ref{app:rq6}), also explains why the most severe retry-loop cases in Appendix~\ref{app:rq6} concentrate in the MCPBench agent stack.

\begin{table}[h]
\centering
\caption{Intrinsic orchestration differences across the three stacks. Items listed as ``LLM decides'' indicate the stack has no built-in mechanism and delegates the decision to the model on the next orchestration step.}
\label{tab:stack_details}
\small
\begin{tabular}{@{}p{3.0cm}p{3.2cm}p{3.2cm}p{3.2cm}@{}}
\toprule
& \textbf{MCPBench} & \textbf{OpenAI (Agents SDK)} & \textbf{LangChain (\texttt{create\_agent})} \\
\midrule
Orchestration pattern & Multi-round plan-and-execute with JSON planner output & Single message loop with native function calling & ReAct state graph on LangGraph (agent/tools nodes) \\
Tool invocation & Structured JSON prompt (\texttt{planned\_tools} list) & Native function calling (\texttt{tool\_choice} required) & Native tool calling on \texttt{ChatOpenAI} \\
Parallel tool calls & Explicit per-round, per-server semaphore & SDK-level fan-out & \texttt{asyncio.gather} in \texttt{ToolNode} \\
Tool-error handling & Error text added to accumulated information buffer & Returned as \texttt{isError} tool result & Caught by middleware, returned as tool message \\
Tool-call retry policy & LLM decides (next round) & LLM decides (next turn) & LLM decides (next agent step) \\
Stop decision & Explicit \texttt{should\_continue} Boolean per round & Implicit (assistant message with no tool calls) & Implicit (AIMessage with no tool calls) \\
Loop bound & 40 rounds & 40 turns & 25 supersteps (LangGraph default) \\
Reasoning format & Free text in JSON \texttt{reasoning} field plus accumulated buffer & Native assistant/tool message history & AI/tool message list on shared state \\
\bottomrule
\end{tabular}
\end{table}

\section{Behavioral Adaptation: Reasoning Traces and Retry Loops}
\label{app:rq6}

\begin{figure}[h]
    \centering
    \includegraphics[width=0.95\linewidth]{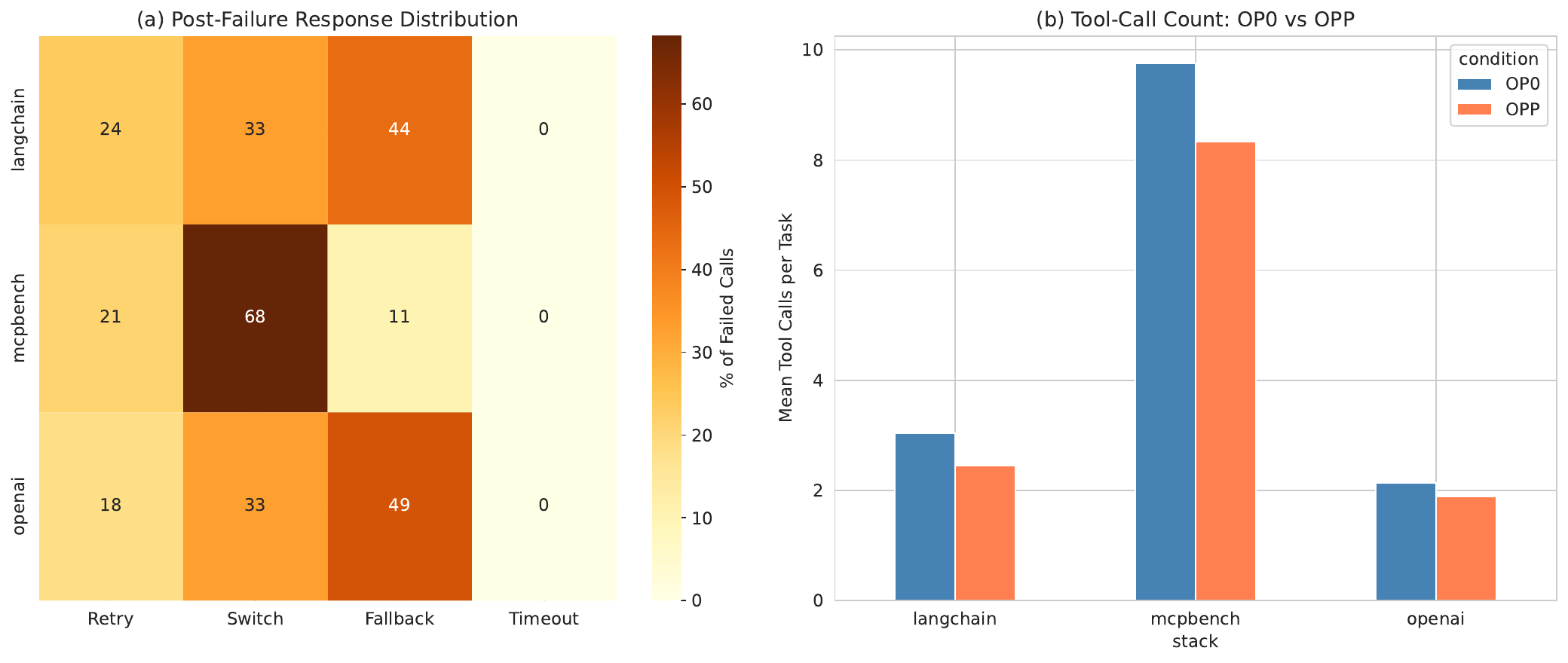}
    \caption{Post-failure behavior by stack. (a) Distribution of the immediate agent response after each fault-adjacent tool call, in percent of failed calls per stack; MCPBench is switch-dominant while LangChain and OpenAI are fallback-leaning. (b) Mean tool calls per task under OP0 versus OPP; within each stack the difference is small and non-significant (paired Wilcoxon, $p \geq 0.09$).}
    \label{fig:rq6_heatmap}
\end{figure}

\paragraph{Fault-response classification.}
For each fault-injected tool call under OPP we classify the agent's immediate response into one of four mutually exclusive actions based on the subsequent tool-call trace: \emph{retry} (the next tool call targets the same tool that just failed), \emph{switch} (the next tool call targets a different tool), \emph{fallback} (no further tool call is issued and the run completes) or \emph{timeout} (no further tool call is issued and the run does not complete).
Across $1{,}302$ fault-adjacent calls the three stacks produce distinct behavioral signatures (Figure~\ref{fig:rq6_heatmap}a).
MCPBench is switch-dominant: 68\% of failures are followed by a call to a different tool, with only 11\% resulting in fallback and 21\% in retry.
LangChain splits 44\% fallback, 33\% switch, 24\% retry.
OpenAI is fallback-leaning: 49\% fallback, 33\% switch, 18\% retry.
No stack records any timeout in this classification: every fault-adjacent call is followed either by another tool call or by a completed run terminating in text.
Pooled across stacks, 57\% of failures are followed by a switch, 22\% by fallback, and 21\% by retry; the pooled picture is dominated by MCPBench simply because its plan-and-execute loop issues far more tool calls per run (mean 8.3 under OPP, versus 2.5 for LangChain and 1.9 for OpenAI; Figure~\ref{fig:rq6_heatmap}b).
A paired Wilcoxon comparison of tool-call counts between OP0 and OPP is non-significant within every stack ($p \geq 0.09$), so the fault injection does not appreciably change how much the agent tries to use tools; it changes \emph{which} tool the agent calls next.

\paragraph{Reasoning trace analysis.}
Examining chain-of-thought traces from MCPBench/120b runs (the only stack that exposes an explicit \texttt{reasoning} field on every planner call) reveals a recurring pattern: the model observes a tool failure, hypothesises a parameter or format issue (``maybe need correct usage'', ``proper parameters''), and either re-issues the same call with a minor variation or pivots to a nearby tool.
Precise fault diagnosis, for instance distinguishing a permanent error from a transient one, is absent in all examined traces (see \texttt{rq6\_reasoning\_excerpts.md} for the curated excerpts).
Agents therefore treat all failures as potentially transient, which is consistent with the switch- and retry-heavy response distribution above.

\paragraph{Retry loops.}
We separately flag runs in which the agent issues three or more consecutive calls to the same failing tool.
Out of 38 such runs across all OPP conditions, 25 occur in MCPBench, 7 in LangChain, and 6 in OpenAI.
The bias toward MCPBench is consistent with its richer LLM-call retry machinery (Appendix~\ref{app:stacks}), which gives the model the most opportunities to re-issue the same tool call before the planner aborts.
The longest chain is \texttt{scientific\_computing\_000} (mcpbench/120b/seed\_3), with 12 consecutive retries over 22 tool calls and 47{,}761 tokens; the most token-expensive chain is the same task on 20b/seed\_1 (52 tool calls, 107{,}849 tokens).

\section{Task-Domain Sensitivity}
\label{app:rq7}

\paragraph{Task structural features.}
We extract two structural features for each of the 25 tasks: \emph{chain depth} (number of distinct tool invocations in the dependency analysis, ranging from 6 to 29) and \emph{domain class} (one of six categories: stateless retrieval, pure computation, multi-step sequential, stateful financial, real-time API, developer tooling).

\begin{figure}[h]
    \centering
    \includegraphics[width=0.95\linewidth]{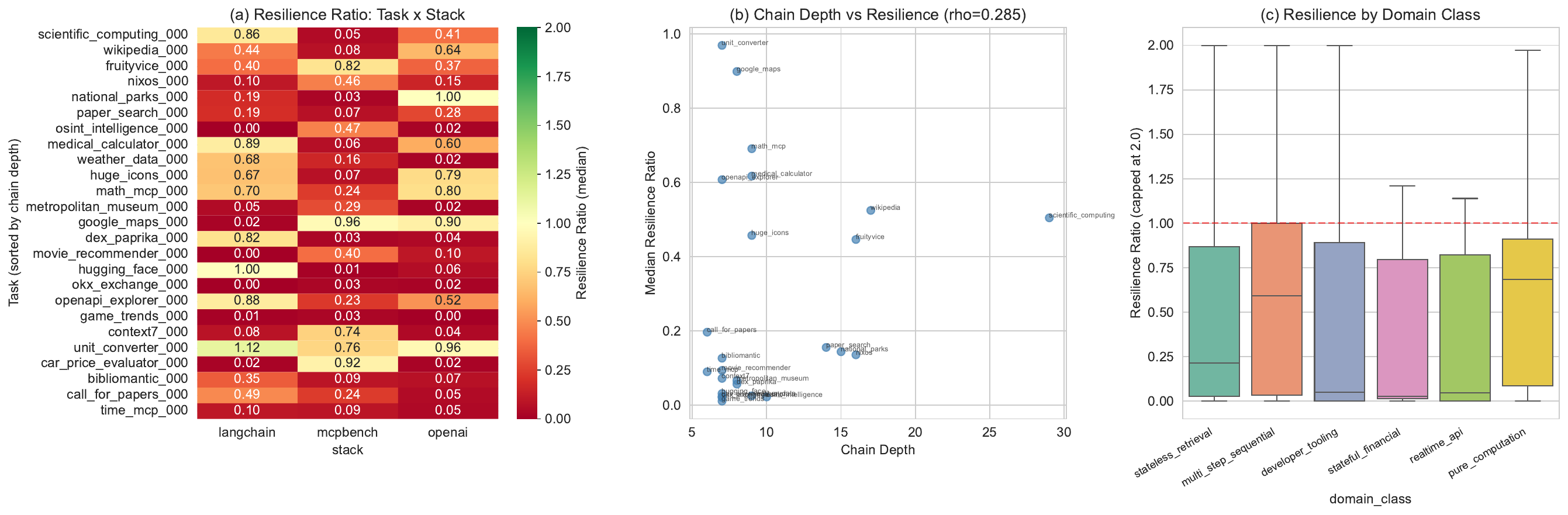}
    \caption{Task-domain sensitivity. (a) Per-task resilience ratio (median of OPP/OP0 across seeds) by stack, with tasks sorted by chain depth. (b) Task-median resilience ratio versus chain depth, labeled by task; the observed Spearman correlation is $+0.29$, $p = 0.17$. (c) Distribution of per-cell resilience ratios (capped at 2.0) by domain class; real-time API, stateful financial, and developer-tooling tasks cluster at the low-resilience end.}
    \label{fig:rq7_heatmap}
\end{figure}

\paragraph{Chain depth and resilience.}
We first examine the bivariate association between chain depth and per-task resilience, where the per-task resilience ratio is the median of OPP/OP0 task-completion scores across stacks, models, and seeds (Figure~\ref{fig:rq7_heatmap}b).
The Spearman correlation is $\rho = 0.29$ ($p = 0.17$, $n = 25$), positive but not significant.
Chain depth, however, is itself correlated with mean fault dose ($\rho = 0.47$, $p = 0.02$), since longer dependency chains present more tool-call opportunities for faults to fire on.
Partialling out fault dose via rank regression leaves a small and non-significant residual association ($\rho_{\text{partial}} = 0.35$, $p = 0.09$).
We therefore do not claim that chain depth causally drives vulnerability in this dataset; we only report that the simple bivariate picture is confounded by fault exposure.

\paragraph{Domain-level comparison.}
Task-median resilience ratios vary substantially across the six domain classes (Figure~\ref{fig:rq7_heatmap}a,c).
Stateful financial tasks (dex\_paprika, okx\_exchange) are the least resilient (median 0.04, $n = 2$), followed closely by developer-tooling tasks (context7, hugging\_face, nixos, osint\_intelligence; median 0.05, $n = 4$) and real-time API tasks (game\_trends, google\_maps, time\_mcp, weather\_data; median 0.06, $n = 4$).
Pure-computation tasks (math\_mcp, scientific\_computing; median 0.60, $n = 2$) and multi-step sequential tasks (median 0.61, $n = 5$) are the most resilient; stateless-retrieval tasks sit in between (median 0.18, $n = 8$).
A Kruskal-Wallis test across the six domain classes is marginal ($H = 9.75$, $p = 0.08$).
With as few as two tasks per class for pure computation and stateful financial, this comparison is underpowered; we report it as descriptive evidence that certain task profiles are reliably harder under OPP, not as a confirmatory domain-level effect.

\section{Reproducibility Details}
\label{app:reproducibility}

\paragraph{Serving hardware and software.}
Both backbones (gpt-oss-20B, gpt-oss-120B) run locally on a single Apple Mac Studio equipped with an M3 Ultra system-on-chip (28-core CPU, 60-core GPU, 32-core Neural Engine, 96\,GB unified memory, 1\,TB SSD), served through LM Studio 0.3.39 at their released MXFP4 precision. No additional quantization is applied, and no per-model prompt tuning is used; cross-model differences in our results are therefore attributable to model scale alone.

\paragraph{Campaign execution history.}
The full campaign underlying the 1{,}350 reported cells was executed over five successive rounds. Each round re-ran cells that had failed in prior rounds due to misconfigurations or timeouts. The first two rounds used a 1{,}500\,s timeout and 20 execution rounds; from the third round on, the limits were raised to the final policy of Section~\ref{sec:experiments-execution}. Of the 1{,}350 reported cells, 875 fault-free (Proxy-off and OP0) cells come from the first two rounds, and all of them completed without reaching the tighter limits (longest run 1{,}449\,s, at most 19 MCPBench rounds). The remaining 25 fault-free cells and all 450 OPP cells ran under the final policy, as did all 51 reported non-completions. Cumulative compute across the five-round campaign, including all re-runs of cells that had failed under tighter timeout or round limits, was approximately 237 wall-clock hours on the Mac Studio. Aggregate compute for the final 1{,}350 reported cells alone was approximately 58 wall-clock hours of agent execution plus about 18 hours of LLM-as-a-judge API time.


\section{Licenses of Used Assets}
\label{app:licenses}

The paper relies on several public assets, all credited in the text and bibliography and used in compliance with their licenses.
\textbf{MCP-Bench}~\cite{wang_mcp-bench_2025} is the benchmark suite (Apache License, Version 2.0); we fork it and carry forward upstream attribution.
The \textbf{OpenAI Agents SDK}~\cite{openai_agent_sdk} and \textbf{LangChain/LangGraph}~\cite{langchain_langgraph} are both distributed under the MIT License.
The \textbf{gpt-oss-20B} and \textbf{gpt-oss-120B} model backbones are released by OpenAI under the Apache License, Version 2.0; they are served locally via LM Studio 0.3.39.
Bundled MCP server implementations under \texttt{mcp\_servers/} retain their per-server upstream licenses (mostly MIT).
The calibration corpus is mined via the GitHub REST API from 1,774 public repositories listed in \emph{MCP Market} and \emph{Awesome MCP Servers}, in compliance with GitHub's terms of service; raw issue and pull-request narratives are deliberately not redistributed, and reviewers can regenerate the raw corpus using their own \texttt{GITHUB\_TOKEN}.
Our own contributions are released under the Apache License, Version 2.0 (source code) and the Creative Commons Attribution 4.0 International license (CC-BY-4.0) for data artifacts (fault profiles, per-task result JSONs, fired-faults logs, and per-thread classifier outputs).
A path-by-path mapping of licenses, the upstream attribution required by Apache 2.0 section 4(d), and a list of pinned commit SHAs and release tags for the third-party code dependencies used in the reported campaigns are documented in the artifact repository (\texttt{LICENSES-README.md}, \texttt{NOTICE}, and \texttt{REPRODUCING.md}).


\newpage
\input{checklist.tex}

\end{document}

%% file: checklist.tex
\section*{NeurIPS Paper Checklist}

\begin{enumerate}

\item {\bf Claims}
    \item[] Question: Do the main claims made in the abstract and introduction accurately reflect the paper's contributions and scope?
    \item[] Answer: \answerYes{}
    \item[] Justification: The abstract and introduction (Section~\ref{sec:introduction}) state three contributions, each supported by a dedicated section: (1)~\injname{}, the protocol-aware, drop-in MCP fault-injection middleware with seedable persistence schedules, is described in Section~\ref{sec:design}; (2)~the MCP fault model, calibrated to the observed incident distribution of the public MCP ecosystem and mined from 57{,}240 threads across 1{,}774 open-source repositories, is detailed in Section~\ref{sec:methodology}; and (3)~the 1{,}350-cell benchmark study across two backbone sizes, three agent stacks, and three fault injection conditions is reported in Section~\ref{sec:results} (with apparatus-validation, mechanistic, and task-level analyses in Appendices~\ref{app:rq1},~\ref{app:rq2},~\ref{app:rq3},~\ref{app:rq6}, and~\ref{app:rq7}), where the quantitative headline findings cited in the abstract (severe score degradation, tool bypass as a baseline behavior that faults expose, no resilience benefit from scale, and suggestive stack-mediated resilience) are each backed by analyses in the main results section and the supporting appendices. Scope boundaries (non-adversarial threat model, MCP tool-call boundary only, two backbone sizes, English-language mining corpus) are stated explicitly across Sections~\ref{sec:background},~\ref{sec:methodology}, and~\ref{sec:experiments} so that the claims reflect what the experiments actually test.

\item {\bf Limitations}
    \item[] Question: Does the paper discuss the limitations of the work performed by the authors?
    \item[] Answer: \answerYes{}
    \item[] Justification: Section~\ref{sec:limitations} is a dedicated Limitations section that discusses construct, internal, and external validity. For construct validity, we acknowledge that the LLM-as-a-judge evaluator is stochastic (mitigated by five evaluations per run with randomized rubric order and median scoring) and rewards some ungrounded answers, that the fault profile mined from GitHub issues and pull requests inherits their reporting bias and is therefore treated as a stress profile, and that the content operators omit schema drift, payload truncation, and partial result omission. For internal validity, we state that the shared 16K-token context window and cross-stack differences in tool-call volume are co-equal alternative explanations for the reported differences, that exposure and error handling remain partially entangled, that the claim that longer tool-call chains amplify exposure is a hypothesis, and that zero-dose cells (12.2\% of OPP cells) and three seeds per cell limit power for per-task comparisons and rare interactions. For external validity, we flag that both backbones come from one model family under one serving setup, that MCP-Bench is easy without faults (OP0 completion 95 to 100\%), that the study is powered for the ecosystem's 91\%-permanent incident mix rather than for transient or intermittent regimes, that the fault profile reflects the open-source MCP ecosystem as of September 2025, and that the stacks and the 25 single-server tasks do not cover multi-agent, human-in-the-loop, multi-server, or longer-horizon settings. The paper rests on three load-bearing assumptions: a non-adversarial threat model and the MCP tool-call boundary as the correct injection point (both stated in Section~\ref{sec:background}), and the representativeness of the mined fault profile (stated in Sections~\ref{sec:methodology} and~\ref{sec:limitations}). If the non-adversarial assumption is violated, an intelligent adversary may exploit patterns (timing side channels, crafted payloads) that our calibrated profile does not capture, so the measured degradation would understate worst-case behavior; if the tool-call boundary is not the dominant actuator for a given deployment, other MCP surfaces (prompts, resources, streaming) may constitute additional and uncharacterized failure sources; if the open-source mining corpus is not representative of production usage, enterprise deployments (with private tools, stricter SLAs, different failure distributions) may experience degradation profiles that differ from those reported here. Computational efficiency is not a central concern for \injname{}, which is a transparent MCP middleware rather than a compute-bound algorithm and scales linearly with the number of tool calls per task (independent of any training-data notion of dataset size); we nonetheless characterize its runtime cost in Appendix~\ref{app:rq1}, where OP0 adds roughly 0.1 to 0.5\,s per tool invocation (latency deltas ranging from $-$12\% to $+$21\% across stacks) with no Bonferroni-significant score difference from proxy-off. Privacy and fairness considerations do not directly apply, as \injname{} operates on synthetic tool-call traffic in non-human-subject benchmark tasks and neither collects user data nor makes decisions about people; the mined corpus draws only on public GitHub issue and pull-request metadata from open-source MCP repositories, with no personal or sensitive content retained. Together these scope boundaries bound the generalizability of specific quantitative findings (e.g., exact degradation percentages), while the qualitative conclusions (no resilience benefit from scale, partial degradation as the most common outcome, tool bypass as a baseline behavior) are likely more robust than the suggestive stack-level differences.

\item {\bf Theory assumptions and proofs}
    \item[] Question: For each theoretical result, does the paper provide the full set of assumptions and a complete (and correct) proof?
    \item[] Answer: \answerNA{}
    \item[] Justification: The paper presents a systems artifact (\injname{}), an empirically mined MCP fault model, and an empirical benchmark study; it does not claim theoretical results, theorems, lemmas, or proofs. The load-bearing assumptions of the empirical study (non-adversarial threat model, MCP tool-call boundary as the injection actuator, representativeness of the mined profile) are stated and discussed in Section~\ref{sec:background} and Section~\ref{sec:limitations}.

    \item {\bf Experimental result reproducibility}
    \item[] Question: Does the paper fully disclose all the information needed to reproduce the main experimental results of the paper to the extent that it affects the main claims and/or conclusions of the paper (regardless of whether the code and data are provided or not)?
    \item[] Answer: \answerYes{}
    \item[] Justification: The paper discloses the full recipe needed to regenerate the main experimental results. Section~\ref{sec:design} describes the \injname{} middleware (interposition model in Figure~\ref{fig:architecture}, four-stage Observe/Decide/Act/Record pipeline, the six fault operators plus the null passthrough in Table~\ref{tab:operators}, the three persistence schedules, scope and composition rules, and the seeded randomness that drives the fault schedules, with every fired fault logged). Section~\ref{sec:methodology} and Appendix~\ref{app:profile-derivation} specify how the OPP fault profile was calibrated (57{,}240 threads across 1{,}774 open-source MCP repositories, classification pipeline, and the 18 injection rules with persistence semantics that constitute the profile). Section~\ref{sec:experiments} then gives the full-factorial design ($2 \times 3 \times 3 \times 3 \times 25 = 1{,}350$ cells), the backbones (gpt-oss-20B and gpt-oss-120B, both publicly available open-weight models served with a 16{,}000-token context and default decoding), the serving hardware (a single Apple Mac Studio with M3 Ultra SoC, 28-core CPU, 60-core GPU, 96\,GB unified memory), the three agent stacks (documented in detail in Appendix~\ref{app:stacks} so that plan format, parallelism, retry behavior, and loop bounds are all made explicit), the 25 single-server MCP-Bench tasks with distraction servers disabled, the per-task timeout and loop bounds governing all reported cells (3{,}000\,s wall-clock; 40 rounds, 40 turns, or 25 supersteps depending on the stack), the judge configuration (gpt-4o-mini-2024-07-18, five evaluations per run with randomized rubric order and median aggregation), the paired within-task comparison $\Delta = \text{OPP} - \text{OP0}$, and the three seeds that fix the benchmark, injector, and framework-level RNGs. Appendices~\ref{app:rq1}, \ref{app:rq2}, \ref{app:rq3}, \ref{app:rq6}, \ref{app:rq7}, and \ref{app:case-studies} report the supplementary analyses, statistical tests, and behavioral traces underlying each headline finding. Because the backbones are open-weight, MCP-Bench is a public benchmark, and every relevant parameter of the injector, run policy, and judge is specified, a reader with the same models, tasks, and fault profile can reconstruct the experimental matrix from the paper alone. Access to the \injname{} middleware, the fault profile JSON files, the campaign scripts, the per-task result JSONs, and the analysis notebooks is addressed in the reply to the next question.

\item {\bf Open access to data and code}
    \item[] Question: Does the paper provide open access to the data and code, with sufficient instructions to faithfully reproduce the main experimental results, as described in supplemental material?
    \item[] Answer: \answerYes{}
    \item[] Justification: The public artifact at \url{https://gitlab.com/uniluxembourg/snt/sedan/papers/mcp-injector}, archived at \url{https://doi.org/10.5281/zenodo.23573765}, contains the \injname{} proxy middleware, the calibrated fault profile (\texttt{fault\_modes/fault\_mode\_1.json}), the campaign and baseline runners, the full 1,350-cell evidence base under \texttt{analysis/results\_formatted/}, the fault-mining pipeline under \texttt{mining/}, and the analysis notebook that produces every figure and table in the paper. The top-level \texttt{README.md} documents environment setup (LM Studio, conda, API keys), a single-cell quickstart, and full-campaign loops; \texttt{REPRODUCING.md} maps each paper figure and table to the notebook cell or script that produced it. Code is released under Apache 2.0 and data artifacts under CC-BY-4.0 (see \texttt{LICENSE}, \texttt{DATA-LICENSE}, and \texttt{LICENSES-README.md}).

\item {\bf Experimental setting/details}
    \item[] Question: Does the paper specify all the training and test details (e.g., data splits, hyperparameters, how they were chosen, type of optimizer) necessary to understand the results?
    \item[] Answer: \answerYes{}
    \item[] Justification: Section~\ref{sec:experiments} specifies the benchmark (25 single-server MCP-Bench tasks with distraction servers disabled), the four-factor full-factorial design (two backbones, three agent stacks, three proxy conditions, three seeds, 25 tasks, 1{,}350 cells), the backbone configuration (gpt-oss-20B and gpt-oss-120B served at MXFP4 precision via LM Studio 0.3.39 on a single M3 Ultra Mac Studio with a 16{,}000-token context window and default decoding), the three proxy conditions (Proxy-off, OP0, OPP with 18 injection rules from the calibrated profile), the paired within-task design $\Delta = \text{OPP} - \text{OP0}$, the run policy (a 3{,}000\,s per-task timeout and a stack-specific loop bound of 40 rounds, 40 turns, or 25 supersteps, with runs exceeding either limit classified as system-level failures), the judge configuration (gpt-4o-mini-2024-07-18, five evaluations per run with randomized rubric order and median aggregation), and the three metric families (task-quality scores, system-level outcomes, latency decomposition). Run-policy and judge hyperparameters follow MCP-Bench~\cite{wang_mcp-bench_2025}, which is the source of the benchmark itself. Appendix~\ref{app:stacks} details stack-level orchestration differences (plan format, parallelism, retry logic, loop bounds) and Appendix~\ref{app:profile-derivation} details the 18-rule OPP profile and how it was calibrated. No model training is performed (open-weight backbones are used off the shelf), so optimizer and training-hyperparameter details are not applicable.

\item {\bf Experiment statistical significance}
    \item[] Question: Does the paper report error bars suitably and correctly defined or other appropriate information about the statistical significance of the experiments?
    \item[] Answer: \answerYes{}
    \item[] Justification: Every main-claim finding is reported with an effect size and a 95\% confidence interval or an explicit significance test, and distributions are displayed as boxplots (median, IQR, whiskers) rather than symmetric error bars because most quantities of interest (bounded completion rates, resilience ratios, long-tailed latencies) are skewed or bounded. Score degradation under fault injection (Section~\ref{sec:rq3} and Appendix~\ref{app:rq3}) uses paired Wilcoxon signed-rank tests across nine metrics with Benjamini-Hochberg correction (all corrected $p < 0.001$; paired drops are reported with 95\% CIs). Proxy non-intrusiveness (Appendix~\ref{app:rq1}) uses paired Wilcoxon tests across 18 (stack, model, metric) combinations with Bonferroni correction (all corrected $p \geq 0.05$, largest effect size $r = 0.40$). Exposure versus dose (Appendix~\ref{app:rq2}) is examined with a linear mixed-effects model \texttt{score $\sim$ fault\_dose + stack + (1\,|\,task\_id)}, reported with 95\% CIs. Stack and model scaling effects (Sections~\ref{sec:rq4} and~\ref{sec:rq5}) use mixed-effects models reported with 95\% CIs (stack $\times$ condition: $\beta = -0.93$ $[-1.86, -0.01]$ for MCPBench and $-0.84$ $[-1.77, 0.08]$ for OpenAI; model size: $\beta_{\text{model}} = -0.009$ $[-0.39, 0.37]$). Task-characteristic analyses (Appendix~\ref{app:rq7}) use Spearman correlations, partial Spearman via rank regression, and a Kruskal-Wallis test across six domain classes. The factors of variability captured by the analyses are made explicit: the paired within-task design $\Delta = \text{OPP} - \text{OP0}$ (Section~\ref{sec:experiments}) removes task-difficulty variance and isolates condition-induced variance, three seeds per cell capture LLM decoding stochasticity, the random intercept \texttt{(1\,|\,task\_id)} in the mixed-effects model captures task-level heterogeneity, and the judge stochasticity is controlled by scoring each run five times with a randomized rubric order and taking the median. The nonparametric tests (Wilcoxon, Spearman, Kruskal-Wallis) make no Normality assumption on the scores; the mixed-effects models assume Normal residuals and Normal random intercepts, which we do not test formally. All $p$-values are computed via standard library functions (SciPy for Wilcoxon, Spearman, and Kruskal-Wallis; statsmodels for the mixed-effects models; the exposure-versus-dose model of Appendix~\ref{app:rq2} was fit with an equivalent REML implementation that reproduces the statsmodels estimates of the stack $\times$ condition model), multiple comparisons are controlled with Bonferroni or Benjamini-Hochberg as appropriate, and effect sizes are reported alongside the $p$-values (matched-pairs rank-biserial $r$ for Wilcoxon, Spearman $\rho$ for correlations, $\beta$ coefficients for the mixed-effects model). Figure and table captions restate the test and correction used (e.g., Table~\ref{tab:itt_degradation} reports Benjamini-Hochberg corrected Wilcoxon tests).

\item {\bf Experiments compute resources}
    \item[] Question: For each experiment, does the paper provide sufficient information on the computer resources (type of compute workers, memory, time of execution) needed to reproduce the experiments?
    \item[] Answer: \answerYes{}
    \item[] Justification: The paper discloses the compute needed to reproduce every reported experiment. All agent and backbone inference runs on a single Apple Mac Studio (M3 Ultra system-on-chip, 28-core CPU, 60-core GPU, 32-core Neural Engine, 96\,GB unified memory, 1\,TB SSD) serving gpt-oss-20B and gpt-oss-120B through LM Studio 0.3.39 at MXFP4 precision with a 16{,}000-token context window (Section~\ref{sec:experiments-factors}). Two external API services are additionally used: the LLM-as-a-judge evaluator (gpt-4o-mini-2024-07-18) is called five times per completed cell via the OpenAI API (Section~\ref{sec:experiments-execution}), and the fault-mining classifier (gpt-4o-2024-08-06) was invoked once to label 57{,}240 threads at a total API cost of approximately USD~\$800 (Section~\ref{sec:methodology}). Per-cell compute is bounded by a 3{,}000\,s per-task wall-clock timeout and a stack-specific loop bound (Section~\ref{sec:experiments-execution}); observed per-cell end-to-end and agent-internal execution times are reported in Appendix~\ref{app:rq1}. The final 1{,}350 reported cells required approximately 58 wall-clock hours of agent execution on the Mac Studio plus about 18 hours of LLM-as-a-judge API time (Section~\ref{sec:experiments-execution}); cumulative Mac Studio compute across the five-round campaign was approximately 237 wall-clock hours. The full research project used more compute than the final reported cells: as disclosed in Section~\ref{sec:experiments-execution}, the campaign was executed over five successive rounds in which earlier-round cells that had failed due to misconfiguration or timeout were re-run under progressively larger timeout and round limits; a small amount of additional compute was spent on pilot tuning of timeouts and proxy configuration, but it is negligible relative to the final campaign.
    
\item {\bf Code of ethics}
    \item[] Question: Does the research conducted in the paper conform, in every respect, with the NeurIPS Code of Ethics \url{https://neurips.cc/public/EthicsGuidelines}?
    \item[] Answer: \answerYes{} 
    \item[] Justification: The authors have reviewed the NeurIPS Code of Ethics and the research conforms to it. The study does not involve human subjects, crowdsourcing, or user interaction, so IRB and compensation requirements do not apply (Questions 14 and 15 answer \answerNA{}). The only external corpus used is public GitHub issue and pull-request metadata from 1{,}774 open-source MCP repositories (57{,}240 threads), from which an aggregate 18-rule fault profile is derived; no personal or sensitive content is retained. Experiments are run on off-the-shelf open-weight backbones (gpt-oss-20B and gpt-oss-120B) with synthetic tool-call traffic on public MCP-Bench tasks, so no personally identifiable information is processed. The released artifacts (the \injname{} middleware, the calibrated fault profile \texttt{fault\_mode\_1.json}, the campaign runners, the analysis notebook, and the 1{,}350-cell evidence base) are published at \url{https://gitlab.com/uniluxembourg/snt/sedan/papers/mcp-injector}, archived at \url{https://doi.org/10.5281/zenodo.23573765}, under Apache 2.0 (code) and CC-BY-4.0 (data), with licenses and intended use documented in \texttt{LICENSE}, \texttt{DATA-LICENSE}, and \texttt{LICENSES-README.md}. The threat model is explicitly non-adversarial (Section~\ref{sec:background}), \injname{} is a transparent local MCP proxy with no credential handling or lateral-movement capability, and the fault profile reproduces naturally occurring open-source failure patterns rather than novel attack payloads, which bounds the dual-use surface. Transparency obligations are met by full disclosure of the experimental recipe, statistical tests with multiple-comparison corrections, both primary (all-runs) and completed-only analyses, and an explicit discussion of construct, internal, and external validity threats (Section~\ref{sec:limitations}). The submitted version was fully anonymized; this camera-ready version is de-anonymized as required.

\item {\bf Broader impacts}
    \item[] Question: Does the paper discuss both potential positive societal impacts and negative societal impacts of the work performed?
    \item[] Answer: \answerYes{}
    \item[] Justification: The paper includes a dedicated Broader impacts section (Section~\ref{sec:broader-impacts}) covering positive impacts (reproducible pre-deployment stress testing of tool-calling agents and a fault profile calibrated to the observed incident distribution of the public MCP ecosystem), risks under intended use (the proxy as a high-privilege surface for tool-call data), miscalibration risks (the profile should be treated as a stress test rather than a certification instrument), and misuse risks (the operator primitives repurposed as an adversarial MCP middleware), together with three mitigations: null-injector default, fault-header tagging of injected responses, and pre-deployment-only framing.
    
\item {\bf Safeguards}
    \item[] Question: Does the paper describe safeguards that have been put in place for responsible release of data or models that have a high risk for misuse (e.g., pre-trained language models, image generators, or scraped datasets)?
    \item[] Answer: \answerYes{}
    \item[] Justification: We release the \injname{} proxy and a text-only empirical fault-mode dataset mined from public GitHub issues and pull requests. We do not release any pre-trained language model, image generator, or scraped image dataset. The proxy is explicitly framed under a non-adversarial threat model (Section~\ref{sec:background}) and is architecturally scoped to MCP deployments the operator already controls (the agent must be reconfigured to route through the proxy, which provides no mechanism to attack third-party servers, harvest credentials, or persist beyond the session). The released artifact includes a usage note restricting use to research and dependability evaluation, logs every fault activation for auditability, and ships aggregated thread metadata and failure-mode classifications rather than the raw scraped corpus.

\item {\bf Licenses for existing assets}
    \item[] Question: Are the creators or original owners of assets (e.g., code, data, models), used in the paper, properly credited and are the license and terms of use explicitly mentioned and properly respected?
    \item[] Answer: \answerYes{}
    \item[] Justification: All external assets used in the paper (the MCP-Bench benchmark, the OpenAI Agents SDK, LangChain/LangGraph, the gpt-oss model backbones, the bundled MCP server implementations, and the GitHub-mined calibration corpus) are credited in the text and bibliography and used in compliance with their licenses. Appendix~\ref{app:licenses} lists each asset together with its license (Apache 2.0, MIT, or CC-BY-4.0) and notes that mining of the calibration corpus respects GitHub's terms of service. The artifact repository ships \texttt{LICENSES-README.md} (path-by-path license map), \texttt{NOTICE} (Apache 2.0 section 4(d) upstream attribution), and \texttt{REPRODUCING.md} (pinned commit SHAs and release tags for the third-party code dependencies used in the reported campaigns).

\item {\bf New assets}
    \item[] Question: Are new assets introduced in the paper well documented and is the documentation provided alongside the assets?
    \item[] Answer: \answerYes{}
    \item[] Justification: The paper releases four new assets, each documented alongside the assets in the public artifact repository \url{https://gitlab.com/uniluxembourg/snt/sedan/papers/mcp-injector} (archived snapshot: \url{https://doi.org/10.5281/zenodo.23573765}): the \injname{} proxy middleware and its wiring in the benchmark runner, the calibrated fault profile \texttt{fault\_modes/fault\_mode\_1.json} (and its null-injector counterpart \texttt{fault\_mode\_0.json}), the per-thread classifier outputs from the 57{,}240-thread / 1{,}774-repository mining pipeline under \texttt{mining/}, and the full 1{,}350-cell campaign result corpus under \texttt{analysis/results\_formatted/} together with the notebook that produces every paper figure and table. Documentation is shipped with the assets: \texttt{mcp-bench/README.md} (artifact summary, environment and API-key setup, LM Studio configuration, quickstart, full-campaign loops, standalone injector launch path, and verification of fired faults), \texttt{mcp-bench/REPRODUCING.md} (per-figure and per-table mapping to the producing notebook cell or script), \texttt{fault\_modes/README.md} (operator and persistence semantics), and \texttt{mining/README.md} (September 2025 snapshot date, scope statement, rate-limit and roughly USD~\$800 OpenAI budget, cold-regeneration recipe, and cross-reference to Section~\ref{sec:limitations}). License metadata is shipped as \texttt{LICENSE} (Apache 2.0 for code), \texttt{DATA-LICENSE} (CC-BY-4.0 for fault profiles, fired-fault logs, per-task result JSONs, and per-thread classifier outputs), \texttt{NOTICE} (upstream attribution to \texttt{accenture/mcp-bench}, and the \injname{} authors for our additions, listed in \texttt{AUTHORS}), and \texttt{LICENSES-README.md} (a path-to-license map); per-file SPDX headers declare the same terms on new sources. No human-subject consent applies: the mined corpus is drawn from public GitHub issue and pull-request threads, and we redistribute only thread identifiers, repository identifiers, classification labels, persistence categories, thread titles, and one-sentence classifier rationales, with raw issue or pull-request bodies, comment chains, and review-comment chains excluded, so third-party repository licenses do not propagate into our CC-BY-4.0 scope. The release carries no development history: it is published as a release snapshot, with author home-directory paths, operational logs, and local editable-install entries removed, and an automated sweep confirms that no credentials, API keys, or local working notes are present in the published tree.

\item {\bf Crowdsourcing and research with human subjects}
    \item[] Question: For crowdsourcing experiments and research with human subjects, does the paper include the full text of instructions given to participants and screenshots, if applicable, as well as details about compensation (if any)?
    \item[] Answer: \answerNA{}
    \item[] Justification: The paper does not involve crowdsourcing or human subjects. All experiments use automated runs of open-weight backbones (gpt-oss-20B and gpt-oss-120B) on public MCP-Bench tasks with an LLM-as-a-judge evaluator (gpt-4o-mini-2024-07-18), and the fault-mode calibration corpus is classified automatically by gpt-4o-2024-08-06 from public GitHub issue and pull-request metadata. The only human labeling is the inter-annotator reliability audit reported in Question~16 and Section~\ref{sec:methodology} (Cohen's $\kappa = 0.93$ for failure mode, $\kappa = 0.80$ for persistence), which was performed by the authors on 117 public GitHub threads as a quality check on the automated classifier; no crowdworkers were recruited or compensated, and no data about the annotators was collected. This is consistent with the \answerNA{} for Question~15 (IRB) and with the Code-of-Ethics declaration in Question~9.

\item {\bf Institutional review board (IRB) approvals or equivalent for research with human subjects}
    \item[] Question: Does the paper describe potential risks incurred by study participants, whether such risks were disclosed to the subjects, and whether Institutional Review Board (IRB) approvals (or an equivalent approval/review based on the requirements of your country or institution) were obtained?
    \item[] Answer: \answerNA{}
    \item[] Justification: The study involves no human subjects, interviews, surveys, or crowdsourcing, so IRB review (or equivalent) was not required. The fault-profile dataset underlying \injname{} is mined programmatically from public open-source MCP repositories (issue and discussion threads), and is analyzed only in aggregate (fault-category proportions and persistence distributions); no personal, private, or identifiable data are collected or released. All dependability experiments are fully automated: agent stacks operate on MCP tool calls, LLM backbones are served locally, and evaluation is performed by an automated judge. No participants were recruited, compensated, or exposed to risk. The inter-annotator reliability audit reported in Question~16 and Section~\ref{sec:methodology} was performed by the authors on 117 public GitHub threads; it does not constitute human-subjects research, and no data about the annotators was collected.

\item {\bf Declaration of LLM usage}
    \item[] Question: Does the paper describe the usage of LLMs if it is an important, original, or non-standard component of the core methods in this research? Note that if the LLM is used only for writing, editing, or formatting purposes and does \emph{not} impact the core methodology, scientific rigor, or originality of the research, declaration is not required.
    \item[] Answer: \answerYes{}
    \item[] Justification: LLMs are a core, non-standard component of this work in three distinct roles, each documented in the paper. (i)~Agent backbones under test: gpt-oss-20B and gpt-oss-120B are the subjects under evaluation, served locally via LM Studio 0.3.39 at MXFP4 precision with a 16{,}000-token context window and default decoding across three agent stacks (MCPBench multi-round plan-and-execute, OpenAI Agents SDK, LangChain/LangGraph), as specified in Section~\ref{sec:experiments-factors}. (ii)~LLM-as-a-judge evaluator: gpt-4o-mini-2024-07-18 scores each completed run five times with randomized rubric order and median aggregation, as specified in Section~\ref{sec:experiments-execution}. (iii)~Fault-mining classifier: gpt-4o-2024-08-06 was invoked (temperature~0, fixed JSON schema) to classify 57{,}240 threads from 1{,}774 open-source MCP repositories, yielding 14{,}829 labeled (failure mode, persistence) failure instances, at a total API cost of approximately USD~\$800, as described in Section~\ref{sec:methodology} and Appendix~\ref{app:profile-derivation}; the 18-rule OPP fault profile (\texttt{fault\_modes/fault\_mode\_1.json}) is derived directly from these classifier outputs and is therefore a load-bearing product of LLM use, with reliability supported by an independent human audit (Cohen's $\kappa = 0.93$ for failure mode, $\kappa = 0.80$ for persistence). Any use of LLMs for writing or editing assistance did not influence the scientific claims, experimental design, statistical analyses, or conclusions, and is excluded from this declaration per the NeurIPS LLM policy.

\end{enumerate}

%% file: chaosllm.bib
@article{natella2016assessing,
  title={Assessing dependability with software fault injection: A survey},
  author={Natella, Roberto and Cotroneo, Domenico and Madeira, Henrique S},
  journal={ACM Computing Surveys (CSUR)},
  volume={48},
  number={3},
  pages={1--55},
  year={2016},
  publisher={ACM New York, NY, USA}
}

@article{wang2024gta,
  title={GTA: a benchmark for general tool agents},
  author={Wang, Jize and Zerun, Ma and Li, Yining and Zhang, Songyang and Chen, Cailian and Chen, Kai and Le, Xinyi},
  journal={Advances in Neural Information Processing Systems},
  volume={37},
  pages={75749--75790},
  year={2024}
}

@inproceedings{seo2014programmers,
  title={Programmers' build errors: a case study (at google)},
  author={Seo, Hyunmin and Sadowski, Caitlin and Elbaum, Sebastian and Aftandilian, Edward and Bowdidge, Robert},
  booktitle={Proceedings of the 36th International Conference on Software Engineering},
  pages={724--734},
  year={2014}
}

@article{landis1977measurement,
  title={The measurement of observer agreement for categorical data},
  author={Landis, J. R. and Koch, G. G.},
  journal={Biometrics},
  volume={33},
  number={1},
  pages={159--174},
  year={1977}
}

@INPROCEEDINGS{11105280,
  author={Ehtesham, Abul and Singh, Aditi and Kumar, Saket},
  booktitle={2025 IEEE World AI IoT Congress (AIIoT)}, 
  title={Enhancing Clinical Decision Support and EHR Insights through LLMs and the Model Context Protocol: An Open-Source MCP-FHIR Framework}, 
  year={2025},
  volume={},
  number={},
  pages={0205-0211},
  doi={10.1109/AIIoT65859.2025.11105280}}

@INPROCEEDINGS{11205601,
  author={Vieira Da Silva, Luis Miguel and Köcher, Aljosha and Gehlhoff, Felix},
  booktitle={2025 IEEE 30th International Conference on Emerging Technologies and Factory Automation (ETFA)}, 
  title={Beyond Formal Semantics for Capabilities and Skills: Model Context Protocol in Manufacturing}, 
  year={2025},
  volume={},
  number={},
  pages={1-4},
  doi={10.1109/ETFA65518.2025.11205601}}

@INPROCEEDINGS{11213046,
  author={Feng, Qi and Zhang, Guowei and Chen, Kaize},
  booktitle={2025 7th International Conference on Internet of Things, Automation and Artificial Intelligence (IoTAAI)}, 
  title={Design and Implementation of a Delivery Robot System for Natural Language Interaction}, 
  year={2025},
  volume={},
  number={},
  pages={739-744},
  doi={10.1109/IoTAAI66837.2025.11213046}}

@inproceedings{gim2025pie,
  title={Pie: A Programmable Serving System for Emerging LLM Applications},
  author={Gim, In and Ma, Zhiyao and Lee, Seung-seob and Zhong, Lin},
  booktitle={Proceedings of the ACM SIGOPS 31st Symposium on Operating Systems Principles},
  pages={415--430},
  year={2025}
}

@inproceedings{luo2023self,
  title={Self-attention and transformers: Driving the evolution of large language models},
  author={Luo, Qing and Zeng, Weizhen and Chen, Manni and Peng, Gang and Yuan, Xiaofeng and Yin, Qian},
  booktitle={2023 IEEE 6th International conference on electronic information and communication technology (ICEICT)},
  pages={401--405},
  year={2023},
  organization={IEEE}
}

@ARTICLE{1335465,
  author={Avizienis, A. and Laprie, J.-C. and Randell, B. and Landwehr, C.},
  journal={IEEE Transactions on Dependable and Secure Computing}, 
  title={Basic concepts and taxonomy of dependable and secure computing}, 
  year={2004},
  volume={1},
  number={1},
  pages={11-33},
  doi={10.1109/TDSC.2004.2}}

@INPROCEEDINGS{10206870,
  author={Castro, Jessica and Laranjeiro, Nuno and Vieira, Marco},
  booktitle={2023 53rd Annual IEEE/IFIP International Conference on Dependable Systems and Networks - Supplemental Volume (DSN-S)}, 
  title={Techniques and Tools for Runtime Security Monitoring and Analysis of Microservices}, 
  year={2023},
  volume={},
  number={},
  pages={191-193},
  doi={10.1109/DSN-S58398.2023.00052}}

@ARTICLE{7436642,
  author={Basiri, Ali and Behnam, Niosha and de Rooij, Ruud and Hochstein, Lorin and Kosewski, Luke and Reynolds, Justin and Rosenthal, Casey},
  journal={IEEE Software}, 
  title={Chaos Engineering}, 
  year={2016},
  volume={33},
  number={3},
  pages={35-41},
  doi={10.1109/MS.2016.60}}

@article{cotroneo_dependability_2021,
	title = {Dependability {Assessment} of the {Android} {OS} through {Fault} {Injection}},
	volume = {70},
	issn = {0018-9529, 1558-1721},
	url = {http://arxiv.org/abs/1912.03490},
	doi = {10.1109/TR.2019.2954384},
	language = {en},
	number = {1},
	urldate = {2025-09-04},
	journal = {IEEE Transactions on Reliability},
	author = {Cotroneo, Domenico and Iannillo, Antonio Ken and Natella, Roberto and Rosiello, Stefano},
	month = mar,
	year = {2021},
	note = {arXiv:1912.03490 [cs]},
	pages = {346--361},
}

@misc{li_api-bank_2023,
	title = {{API}-{Bank}: {A} {Comprehensive} {Benchmark} for {Tool}-{Augmented} {LLMs}},
	shorttitle = {{API}-{Bank}},
	url = {http://arxiv.org/abs/2304.08244},
	doi = {10.48550/arXiv.2304.08244},
	language = {en},
	urldate = {2025-09-04},
	publisher = {arXiv},
	author = {Li, Minghao and Zhao, Yingxiu and Yu, Bowen and Song, Feifan and Li, Hangyu and Yu, Haiyang and Li, Zhoujun and Huang, Fei and Li, Yongbin},
	month = oct,
	year = {2023},
	note = {arXiv:2304.08244 [cs]},
}

@misc{xu_tool_2023,
	title = {On the {Tool} {Manipulation} {Capability} of {Open}-source {Large} {Language} {Models}},
	url = {http://arxiv.org/abs/2305.16504},
	doi = {10.48550/arXiv.2305.16504},
	language = {en},
	urldate = {2025-09-04},
	publisher = {arXiv},
	author = {Xu, Qiantong and Hong, Fenglu and Li, Bo and Hu, Changran and Chen, Zhengyu and Zhang, Jian},
	month = may,
	year = {2023},
	note = {arXiv:2305.16504 [cs]},
}

@misc{ruan_identifying_2024,
	title = {Identifying the {Risks} of {LM} {Agents} with an {LM}-{Emulated} {Sandbox}},
	url = {http://arxiv.org/abs/2309.15817},
	doi = {10.48550/arXiv.2309.15817},
	language = {en},
	urldate = {2025-09-04},
	publisher = {arXiv},
	author = {Ruan, Yangjun and Dong, Honghua and Wang, Andrew and Pitis, Silviu and Zhou, Yongchao and Ba, Jimmy and Dubois, Yann and Maddison, Chris J. and Hashimoto, Tatsunori},
	month = may,
	year = {2024},
	note = {arXiv:2309.15817 [cs]},
}

@misc{mialon_gaia_2023,
	title = {{GAIA}: a benchmark for {General} {AI} {Assistants}},
	shorttitle = {{GAIA}},
	url = {http://arxiv.org/abs/2311.12983},
	doi = {10.48550/arXiv.2311.12983},
	language = {en},
	urldate = {2025-09-04},
	publisher = {arXiv},
	author = {Mialon, Grégoire and Fourrier, Clémentine and Swift, Craig and Wolf, Thomas and LeCun, Yann and Scialom, Thomas},
	month = nov,
	year = {2023},
	note = {arXiv:2311.12983 [cs]},
}

@article{yao2024tau,
  title={\(\tau\)-bench: A Benchmark for Tool-Agent-User Interaction in Real-World Domains},
  author={Yao, Shunyu and Shinn, Noah and Razavi, Pedram and Narasimhan, Karthik},
  journal={arXiv preprint arXiv:2406.12045},
  year={2024}
}

@misc{wang_mcp-bench_2025,
	title = {{MCP}-{Bench}: {Benchmarking} {Tool}-{Using} {LLM} {Agents} with {Complex} {Real}-{World} {Tasks} via {MCP} {Servers}},
	shorttitle = {{MCP}-{Bench}},
	url = {http://arxiv.org/abs/2508.20453},
	doi = {10.48550/arXiv.2508.20453},
	language = {en},
	urldate = {2025-09-04},
	publisher = {arXiv},
	author = {Wang, Zhenting and Chang, Qi and Patel, Hemani and Biju, Shashank and Wu, Cheng-En and Liu, Quan and Ding, Aolin and Rezazadeh, Alireza and Shah, Ankit and Bao, Yujia and Siow, Eugene},
	month = aug,
	year = {2025},
	note = {arXiv:2508.20453 [cs]},
}

@misc{openai_agent_sdk,
  author = {{OpenAI}},
  title = {OpenAI Agents SDK},
  year = {2025},
  howpublished = {\url{https://github.com/openai/openai-agents-python}},
  note = {Accessed: 2026-04-13},
}

@misc{langchain_langgraph,
  author = {{LangChain}},
  title = {LangGraph: Build Resilient Language Agents as Graphs},
  year = {2024},
  howpublished = {\url{https://github.com/langchain-ai/langgraph}},
  note = {Accessed: 2026-04-13},
}

@inproceedings{shinn2023reflexion,
  author = {Noah Shinn and Federico Cassano and Beck Labash and Ashwin Gopinath and Karthik Narasimhan and Shunyu Yao},
  title = {Reflexion: Language Agents with Verbal Reinforcement Learning},
  booktitle = {Advances in Neural Information Processing Systems},
  year = {2023},
}

@inproceedings{chen2025acebench,
  author = {Chen Chen and Xinlong Hao and Weiwen Liu and Xu Huang and Xingshan Zeng and Shuai Yu and Dexun Li and others},
  title = {{ACEBench}: A Comprehensive Evaluation of {LLM} Tool Usage},
  booktitle = {Findings of the Conference on Empirical Methods in Natural Language Processing},
  year = {2025},
}

@inproceedings{xiong2025butterfly,
  author = {Qian Xiong and Yuekai Huang and Ziyou Jiang and Zhiyuan Chang and Yujia Zheng and Tianhao Li and Mingyang Li},
  title = {Butterfly Effects in Toolchains: A Comprehensive Analysis of Failed Parameter Filling in {LLM} Tool-Agent Systems},
  booktitle = {Proceedings of the Conference on Empirical Methods in Natural Language Processing},
  year = {2025},
}

@article{yu2025fit_microservices,
  author = {Senyao Yu and Huayao Wu and Xintao Niu and Changhai Nie},
  title = {A Systematic Literature Review on Fault Injection Testing of Microservice Systems},
  journal = {IEEE Transactions on Services Computing},
  year = {2025},
}

@inproceedings{owotogbe2025chaos_llm,
  author = {Joshua Owotogbe},
  title = {Assessing and Enhancing the Robustness of {LLM}-Based Multi-Agent Systems Through Chaos Engineering},
  booktitle = {IEEE/ACM International Conference on AI Engineering (CAIN)},
  year = {2025},
}

@inproceedings{kikuta2025chaoseater,
  author = {Daisuke Kikuta and Hiroki Ikeuchi and Kengo Tajiri},
  title = {{LLM}-Powered Fully Automated Chaos Engineering: Towards Enabling Anyone to Build Resilient Software Systems at Low Cost},
  booktitle = {International Conference on Automated Software Engineering},
  year = {2025},
}

@article{hou2025mcp_landscape,
  author = {Xinyi Hou and Yanjie Zhao and Shenao Wang and Haoyu Wang},
  title = {Model Context Protocol ({MCP}): Landscape, Security Threats, and Future Research Directions},
  journal = {ACM Transactions on Software Engineering and Methodology},
  year = {2025},
}

@article{gupta2026reliabilitybench,
  author = {Aayush Gupta},
  title = {{ReliabilityBench}: Evaluating {LLM} Agent Reliability Under Production-Like Stress Conditions},
  journal = {arXiv preprint arXiv:2601.06112},
  year = {2026},
}

@inproceedings{zhang2024agentsecuritybench,
  author = {Hanrong Zhang and Jingyuan Huang and Kai Mei and Yifei Yao and Zhenting Wang and Chenlu Zhan and Hongwei Wang and Yongfeng Zhang},
  title = {Agent Security Bench ({ASB}): Formalizing and Benchmarking Attacks and Defenses in {LLM}-based Agents},
  booktitle = {International Conference on Learning Representations},
  year = {2024},
  note = {arXiv:2410.02644},
}

@misc{liu2025mcpagentbench,
  author = {Wenrui Liu and Zixiang Liu and Elsie Dai and Wenhan Yu and Lei Yu and Tong Yang},
  title = {{MCPAgentBench}: A Real-world Task Benchmark for Evaluating {LLM} Agent {MCP} Tool Use},
  journal = {arXiv preprint arXiv:2512.24565},
  year = {2025},
}

@article{huang2026agents_fail,
  author = {Donghao Huang and Gauri Malwe and Zhaoxia Wang},
  title = {When Agents Fail to Act: A Diagnostic Framework for Tool Invocation Reliability in Multi-Agent {LLM} Systems},
  journal = {arXiv preprint arXiv:2601.16280},
  year = {2026},
}

@INPROCEEDINGS{grottke2010empirical,
  author={Grottke, Michael and Nikora, Allen P. and Trivedi, Kishor S.},
  booktitle={2010 IEEE/IFIP International Conference on Dependable Systems \& Networks (DSN)},
  title={An empirical investigation of fault types in space mission system software},
  year={2010},
  pages={447-456},
  doi={10.1109/DSN.2010.5544284}}
